\def\RELEASE{1}  %
\def\ANON{0}     %
\def\SQUEEZE{0}  %

\documentclass[sigconf,10pt,screen]{acmart}
\PassOptionsToPackage{hyphens}{url}\usepackage{hyperref}

\usepackage[noend]{algpseudocode}
\usepackage{algorithmicx}
\usepackage{booktabs}
\usepackage{caption}
\usepackage{comment}
\usepackage[inline]{enumitem}
\usepackage{graphicx}
\usepackage{listings}
\usepackage{multirow}
\usepackage{xcolor}
\usepackage{xspace}
\usepackage{libertine}
\usepackage{microtype}
\usepackage{tikz}
\usepackage{paralist}
\usepackage{gnuplot-lua-tikz}

\microtypecontext{spacing=nonfrench}

\hypersetup{
  pdflang={en},
  pdfdisplaydoctitle}

\definecolor[named]{OurPurple}{cmyk}{0.55,1,0,0.15}
\definecolor[named]{OurDarkBlue}{cmyk}{1,0.58,0,0.21}
\hypersetup{
  colorlinks,
  linkcolor=OurPurple,
  citecolor=OurPurple,
  urlcolor=OurDarkBlue,
  filecolor=OurDarkBlue}

\if \SQUEEZE 1
\setlist[itemize]{
  leftmargin=*,
  itemsep=2pt,
  topsep=2pt}

\fi
\def\Snospace~{\S{}}

\if \RELEASE 1
  \def\NOTES{0}
\else
  \def\NOTES{1}
\fi
\if \NOTES 1
  \newcommand{\XXX}[1]{{\color{red}{XXX {#1}}}}
  \newcommand{\matheus}[1]{{\color{violet}{[\textbf{MS:} {#1}]}}}
  \newcommand{\antoine}[1]{{\color{teal}{[\textbf{AK:} {#1}]}}}
  \newcommand{\simon}[1]{{\color{red}{[\textbf{SP:} {#1}]}}}
  \newcommand{\todo}[1]{{\color{blue}{TODO: {#1}}}}
\else
  \newcommand{\XXX}[1]{}
  \newcommand{\matheus}[1]{}
  \newcommand{\antoine}[1]{}
  \newcommand{\simon}[1]{}
  \newcommand{\todo}[1]{}
\fi

\if \ANON 1

\else

\fi

\lstdefinelanguage{C}{
  alsoletter={_},
  morekeywords={
    proto_t,auto,break,case,char,const,continue,default,do,double,else,enum,extern,
    float,for,goto,if,inline,int,long,register,restrict,return,short,signed,
    sizeof,static,struct,switch,typedef,union,unsigned,void,volatile,while,
    _Alignas,_Alignof,_Atomic,_Bool,_Complex,_Generic,_Imaginary,_Noreturn,
    _Static_assert,_Thread_local,sentry_t,qentry_t,proto_t,sched_t,
  },
  sensitive=true,
  morecomment=[l]{//},
  morecomment=[s]{/*}{*/},
  morestring=[b]",
}

\definecolor{myblue}{HTML}{4870dd}
\definecolor{mypink}{HTML}{c02d71}
\lstMakeShortInline[columns=fixed,basicstyle=\ttfamily]|

\acmSubmissionID{48}

\begin{document}
\date{}
\if \ANON 1
  \title{Idle Cycles Are Not Wasted Cycles}
\else
  \title{Idle Cycles Are Not Wasted Cycles}
\fi

\if \ANON 1
  \author{Anonymous Submission \#48 (\pageref{page:last})}
\else
  \author{Matheus Stolet}
  \orcid{0009-0009-0175-4828}
  \email{mstolet@mpi-sws.org}
  \affiliation{
    \institution{Max Planck Institute for Software Systems}
    \country{Germany}
  }
  \author{Simon Peter}
  \orcid{0009-0007-4748-8524}
  \email{simpeter@cs.washington.edu}
  \affiliation{
    \institution{University of Washington}
    \country{USA}
  }
  \author{Antoine Kaufmann}
  \orcid{0000-0002-6355-2772}
  \email{antoinek@mpi-sws.org}
  \affiliation{
    \institution{Max Planck Institute for Software Systems}
    \country{Germany}
  }
\fi

\copyrightyear{2026}
\acmYear{2026}
\setcopyright{cc}
\setcctype{by}
\acmConference[HotNets '26]{The 25th ACM Workshop on Hot Topics in Networks}{November 16--17, 2026}{Salt Lake City, UT, USA}
\acmBooktitle{The 25th ACM Workshop on Hot Topics in Networks (HotNets '26), November 16--17, 2026, Salt Lake City, UT, USA}
\acmDOI{10.1145/3847290.3859941}
\acmISBN{979-8-4007-3029-0/2026/11}

\begin{abstract}
Idle network service cores are treated as wasted compute. This
assumption motivates increasingly sophisticated mechanisms
that reclaim idle cores at microsecond timescales. We argue that
this view no longer matches modern server hardware. On
contemporary multicore processors, active cores compete for a
shared package-level power and thermal budget. Once that budget
becomes the limiting resource, an idle core that waits
efficiently returns compute capacity that hardware can
redistribute to productive work. Measurements on a recent AMD
EPYC processor show how waiting strategy, processor topology,
and idle duration determine this trade-off. Our results suggest
that reclaiming idle cores often yields less benefit than
commonly assumed while introducing substantial scheduling
complexity. We propose a budget-centric view of service-core
systems in which power, rather than core occupancy, becomes the
fundamental resource and waiting policy becomes a first-class
systems design choice.
\end{abstract} 


\maketitle

\section{Introduction}

High-performance network stacks increasingly use
\emph{service-core architectures}: packet processing runs on dedicated
cores as a shared service, and applications communicate via
shared-memory queues rather than system 
calls~\cite{kaufmann:tas,marty:snap,stolet:virtuoso,stolet:chamelio,niu:netkernel}.
Dedicating cores keeps applications and the stack from competing for
caches, branch predictors, and other core-local state; parallelises
processing; removes context switches and privilege transitions from
the data path; and lets one shared instance multiplex many
applications.

But dedicated service cores are rarely busy every cycle. Load
fluctuates at microsecond timescales, and demand does not align
with integer multiples of a core. The community treats these idle cycles as
waste to be eliminated, and has built sophisticated machinery to do so,
such as core reallocation at microsecond granularity, dedicated scheduler
cores, and elaborate handoff 
protocols~\cite{ousterhout:shenango,fried:caladan,marty:snap}.
This machinery follows a seemingly obvious mental model: a core-cycle spent idle
is a core-cycle of computation lost.
We argue that this model is wrong on modern servers, leading
to systems more complex and \emph{less} efficient than doing nothing.

Large multicore processors face fixed power and
thermal budgets~\cite{esmaeilzadeh:darksilicon,hardavellas:darkservers}.
For instance, an AMD EPYC~9655 guarantees 2.6\,GHz across its 
96 cores but boosts a single
core to 4.5\,GHz. 
This gap shows that $N$ cores
do not deliver $N$ cores' worth of peak compute.
Production fleets oversubscribe power, so saturated machines routinely run
against their caps~\cite{fan:power_provisioning,wu:dynamo,li:thunderbolt,kumbhare:power_oversubscription}.
On our
EPYC~9655, a cache-resident parallel workload running on all 96 cores
achieves only 47\% of the per-core throughput it achieves alone.
Thus cores are not independent units of compute: performance depends
on a shared budget.

Below saturation, reclaiming a service core offers little.
If an application is not fully utilising its current cores,
reallocation cannot improve performance. If it is saturating
them, the scenario that motivates microsecond core
reallocation, the processor is constrained by its power
budget. Elastic allocation therefore only helps workloads for
which this underlying accounting no longer holds. 
In that regime, the arithmetic
of idleness inverts and an idle core does not squander its
capacity. Instead the processor's firmware redistributes it in
microseconds with no software in the loop.

We quantitatively characterise this regime and measure how waiting
strategy, idle patterns, placement, and reallocation determine how
much budget idle service cores return. Our
insights lead us to propose a replacement mental model: \textbf{the
  processor is a fixed compute budget, not a collection of cores.}
Below the power cap, cores behave classically; above it, they compete
for a shared budget, and one must consider how much budget they consume
instead of how many cores a service holds. 
How a core \emph{waits}
becomes as important as whether it is allocated: hardware waiting
mechanisms---\texttt{PAUSE}, user-level monitor/wait, halt-class deep
idle---return progressively more budget at higher wakeup
latency without giving up the core. Blocking, by contrast,
releases the core to a scheduler, paying wakeups, cache refills, and
engineering complexity to redistribute a budget the hardware already
redistributes for free. This cost is worthwhile only once idle
periods amortise it; power capping can shift this break-even point
toward longer idle periods.

We distil four lessons for service-core architectures and related
multicore systems.
(1) Waiting returns power and thermal headroom, so idleness is a transfer.
(2) The compute budget is distributed hierarchically across the CPU, so task
placement decides who shares it.
(3) Waiting trades wakeup latency for returned budget,
so waiting policies should adapt to the workload. 
(4) Core reallocation is worthwhile only after blocking overheads
are amortised, so our systems should stop paying the cost of complexity to
eliminate idle cycles that were never wasted to begin with.
\section{Background}%
\label{sec:bg}

Modern network stacks and processors have evolved
considerably over the past decade, changing the trade-offs around dedicated
polling. Many of the assumptions that shaped earlier network stack designs no
longer hold on modern servers.

\subsection{Managing Service Cores}
\label{subsec:bg:service}

Service cores isolate operating system services on dedicated cores to
reduce interference with applications. This separation helps
latency-sensitive systems but complicates allocation. Service demand
varies over time, yet cores are often provisioned for peak load to
maintain low latency during bursts. This leaves reserved capacity idle
when demand dips.

\textbf{\textit{Microsecond core reallocation.}}
Recent work treats idle service cores as stranded
resources that should be reclaimed. These systems dynamically
reallocate cores between applications and services as demand
changes, often at microsecond timescales, to recover otherwise
idle polling cycles~\cite{ousterhout:shenango,
fried:caladan,mcclure:efficient-scheduling}. These mechanisms incur direct
coordination and migration costs and indirect costs from disrupting caches
and other microarchitectural state when a core runs different work.
Both must be amortised over the reclaimed work.

\textbf{\textit{Low-overhead core handoffs.}}
A complementary line of work reduces the latency of handing a
core new work. These systems employ low-overhead
preemption, yielding, and interrupt delivery to shorten the
critical path between an idle core and useful
execution~\cite{kaffes:shinjuku,iyer:concord,
aydogmus:xui,lin:fast_core_scheduling}. 
These mechanisms reduce the cost of scheduling
decisions, but they do not assess if reclaiming idle
service cores is beneficial.

\textbf{\textit{Waiting strategies.}}
Service cores use different waiting strategies for short-term idling.
Busy polling does not reduce core power consumption
but resumes immediately. \texttt{PAUSE} spin-loop hints
let the core reduce power and, with multithreading, enable the peer
thread to use more cycles, while preserving 10--100\,ns-scale
wakeups~\cite{spec:intel_64_ia32,kuns:useridle}. 
\texttt{UMWAIT} and \texttt{MWAIT} save more power by entering deeper idle
states~\cite{spec:intel_64_ia32} but with increased wakeup latency. 
Deeper power-saving states, such as \texttt{HLT}, require slow, interrupt-driven
wakeups. 
These strategies form a spectrum, trading energy and multithreading
efficiency against wakeup latency.

\subsection{Power and Thermal Management in Large Multicore Machines}
\label{subsec:bg:chiplets}

The large multicore processors on which 
these services run operate within a package-level power limit.
Hardware and firmware continuously allocate this finite limit across active
cores, reducing their frequency when concurrent activity
exceeds the package's power or thermal envelope. Thus, activating another core
reduces the frequency and the performance of work running
elsewhere on the socket~\cite{esmaeilzadeh:darksilicon,hardavellas:darkservers}.

\textbf{\textit{Processor performance depends on package activity.}}
We use the AMD EPYC~9655 to illustrate this constraint.
The processor integrates 96 cores within a 400\,W default TDP and advertises 
a maximum single-core boost frequency of 4.5\,GHz, but an all-core boost frequency of
4.1\,GHz~\cite{product:amd_epyc9655}. The gap between maximum single-core and 
all-core frequency exemplifies that the frequency available 
to each core depends on activity across the package.
Consequently, cores devoted to polling can consume power headroom
even when they perform no useful network processing.

\begin{figure}%
\centering
\includegraphics[width=0.46\textwidth]{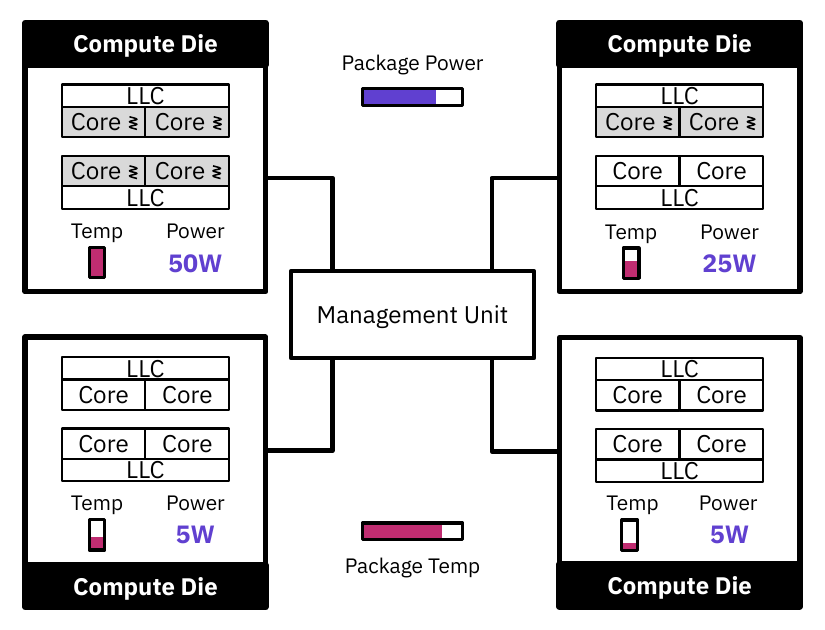}%
\caption{Chiplet-based CPU with four compute
  dies and four cores per die. The management unit controls power
  and frequency across cores by monitoring temperature,
  voltage, and performance counters in individual chiplets
  and keeps the package under its limits.}%
\label{fig:bg:chiplet}%
\Description{Four compute dies. Four cores per die. Central management unit.}%
\end{figure}

\textbf{\textit{Topology influences power and thermal management.}}
The EPYC~9655 organises its cores into twelve eight-core compute dies.
Because each compute die is physically distinct, its power density, temperature,
and available frequency differ as load is distributed across the
package~\cite{ma:tap25d,romano:diffchip}.
A polling thread may therefore interfere differently with application threads
depending on whether they share a core, a die, or the socket. Task
placement is consequently a power-management and
cache-locality decision. \autoref{fig:bg:chiplet} illustrates power
and thermal management in a modern chiplet CPU.

Power constraints have tightened since Dennard scaling ended and are
likely to intensify as core counts grow faster than power efficiency
improves~\cite{esmaeilzadeh:darksilicon,hardavellas:darkservers}.
\section{The Fixed Compute Budget Model}
\label{sec:motiv}

The conventional view of multicore processors treats each core as an
independent unit of compute. Under this model, an idle polling thread
wastes compute because its cycles could instead execute application
instructions. This intuition holds while the processor has 
unused power and thermal headroom. In this section, we evaluate
application performance under constrained thermal and power regimes
with systematic microbenchmarks and show that this intuition breaks
when the processor reaches its package limits.
From these results, we propose lessons
for service-core network stacks (L1--L4) and a new mental model where
the processor behaves like a unit under a fixed compute budget,
rather than a collection of independent cores.
Finally, we use these insights to demonstrate improved end-to-end RPC
throughput with a real service-core network stack.

\subsection{Experimental Setup}

The microbenchmarks are run on a machine with
Linux kernel version 6.12.86 and a single-socket
AMD EPYC~9655 processor with 12 dies, 8 physical cores per die, 96
physical cores in total, and 192 threads with SMT enabled.
We sample package power with Linux RAPL counters and
per-workload CPU frequency with per-core frequency counters.

\textbf{\textit{Benchmarks.}}
Across experiments, we use a matrix multiplication benchmark as
the compute baseline and several polling microbenchmarks as waiting
baselines. The matrix multiplication workload operates on $16\times16$ matrices
with either scalar or 512-bit vector AVX instructions. Matrices fit in the
L1 cache to avoid cache and memory contention.
The polling baseline runs in different modes that
represent different waiting strategies (\texttt{Busy}, \texttt{Pause}, \texttt{Block})
and pseudo-packet processing tasks (\texttt{Work}).
\texttt{Busy} repeatedly spins on
a shared value, \texttt{Pause} performs the same spin loop but
inserts the \texttt{PAUSE} instruction in the poll loop, 
and \texttt{Block} sleeps between polls. 
\texttt{Work} combines polling with simulated IP/TCP header accesses
and checksum computation.
We pin workloads to specific cores and average results over
three 60-second runs, unless otherwise stated.

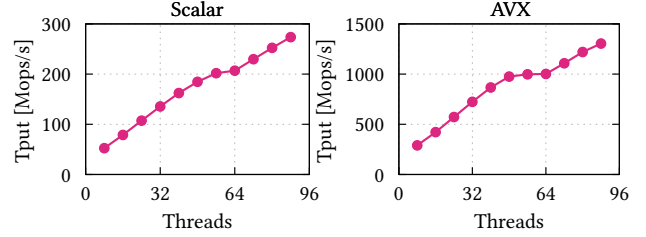
\begin{figure}%
\centering%
\begin{tikzpicture}[gnuplot]
\tikzset{every node/.append style={font={\fontsize{8.0pt}{9.6pt}\selectfont}}}
\path (0.000,0.000) rectangle (8.458,3.378);
\gpcolor{color=gp lt color axes}
\gpsetlinetype{gp lt axes}
\gpsetdashtype{gp dt axes}
\gpsetlinewidth{0.80}
\draw[gp path] (1.057,0.675)--(4.017,0.675);
\gpcolor{color=gp lt color border}
\gpsetlinetype{gp lt border}
\gpsetdashtype{gp dt solid}
\gpsetlinewidth{2.00}
\draw[gp path] (1.057,0.675)--(1.147,0.675);
\draw[gp path] (4.017,0.675)--(3.927,0.675);
\node[gp node right] at (0.983,0.675) {$0$};
\gpcolor{color=gp lt color axes}
\gpsetlinetype{gp lt axes}
\gpsetdashtype{gp dt axes}
\gpsetlinewidth{0.80}
\draw[gp path] (1.057,1.339)--(4.017,1.339);
\gpcolor{color=gp lt color border}
\gpsetlinetype{gp lt border}
\gpsetdashtype{gp dt solid}
\gpsetlinewidth{2.00}
\draw[gp path] (1.057,1.339)--(1.147,1.339);
\draw[gp path] (4.017,1.339)--(3.927,1.339);
\node[gp node right] at (0.983,1.339) {$100$};
\gpcolor{color=gp lt color axes}
\gpsetlinetype{gp lt axes}
\gpsetdashtype{gp dt axes}
\gpsetlinewidth{0.80}
\draw[gp path] (1.057,2.003)--(4.017,2.003);
\gpcolor{color=gp lt color border}
\gpsetlinetype{gp lt border}
\gpsetdashtype{gp dt solid}
\gpsetlinewidth{2.00}
\draw[gp path] (1.057,2.003)--(1.147,2.003);
\draw[gp path] (4.017,2.003)--(3.927,2.003);
\node[gp node right] at (0.983,2.003) {$200$};
\gpcolor{color=gp lt color axes}
\gpsetlinetype{gp lt axes}
\gpsetdashtype{gp dt axes}
\gpsetlinewidth{0.80}
\draw[gp path] (1.057,2.667)--(4.017,2.667);
\gpcolor{color=gp lt color border}
\gpsetlinetype{gp lt border}
\gpsetdashtype{gp dt solid}
\gpsetlinewidth{2.00}
\draw[gp path] (1.057,2.667)--(1.147,2.667);
\draw[gp path] (4.017,2.667)--(3.927,2.667);
\node[gp node right] at (0.983,2.667) {$300$};
\gpcolor{color=gp lt color axes}
\gpsetlinetype{gp lt axes}
\gpsetdashtype{gp dt axes}
\gpsetlinewidth{0.80}
\draw[gp path] (1.057,0.675)--(1.057,2.667);
\gpcolor{color=gp lt color border}
\gpsetlinetype{gp lt border}
\gpsetdashtype{gp dt solid}
\gpsetlinewidth{2.00}
\draw[gp path] (1.057,0.675)--(1.057,0.765);
\node[gp node center] at (1.057,0.429) {$0$};
\gpcolor{color=gp lt color axes}
\gpsetlinetype{gp lt axes}
\gpsetdashtype{gp dt axes}
\gpsetlinewidth{0.80}
\draw[gp path] (2.044,0.675)--(2.044,2.667);
\gpcolor{color=gp lt color border}
\gpsetlinetype{gp lt border}
\gpsetdashtype{gp dt solid}
\gpsetlinewidth{2.00}
\draw[gp path] (2.044,0.675)--(2.044,0.765);
\node[gp node center] at (2.044,0.429) {$32$};
\gpcolor{color=gp lt color axes}
\gpsetlinetype{gp lt axes}
\gpsetdashtype{gp dt axes}
\gpsetlinewidth{0.80}
\draw[gp path] (3.030,0.675)--(3.030,2.667);
\gpcolor{color=gp lt color border}
\gpsetlinetype{gp lt border}
\gpsetdashtype{gp dt solid}
\gpsetlinewidth{2.00}
\draw[gp path] (3.030,0.675)--(3.030,0.765);
\node[gp node center] at (3.030,0.429) {$64$};
\gpcolor{color=gp lt color axes}
\gpsetlinetype{gp lt axes}
\gpsetdashtype{gp dt axes}
\gpsetlinewidth{0.80}
\draw[gp path] (4.017,0.675)--(4.017,2.667);
\gpcolor{color=gp lt color border}
\gpsetlinetype{gp lt border}
\gpsetdashtype{gp dt solid}
\gpsetlinewidth{2.00}
\draw[gp path] (4.017,0.675)--(4.017,0.765);
\node[gp node center] at (4.017,0.429) {$96$};
\draw[gp path] (1.057,2.667)--(1.057,0.675)--(4.017,0.675)--(4.017,2.667)--cycle;
\gpcolor{rgb color={0.863,0.149,0.498}}
\gpsetlinewidth{1.00}
\draw[gp path] (1.304,1.021)--(1.304,1.026);
\draw[gp path] (1.550,1.194)--(1.550,1.203);
\draw[gp path] (1.797,1.386)--(1.797,1.388);
\draw[gp path] (2.044,1.570)--(2.044,1.581);
\draw[gp path] (2.290,1.748)--(2.290,1.755);
\draw[gp path] (2.537,1.897)--(2.537,1.905);
\draw[gp path] (3.030,2.046)--(3.030,2.049);
\draw[gp path] (3.277,2.192)--(3.277,2.208);
\draw[gp path] (3.524,2.342)--(3.524,2.359);
\draw[gp path] (3.770,2.489)--(3.770,2.496);
\gpcolor{color=gpbgfillcolor}
\gpsetpointsize{4.00}
\gp3point{gp mark 7}{}{(1.304,1.023)}
\gpcolor{rgb color={0.863,0.149,0.498}}
\gp3point{gp mark 7}{}{(1.304,1.023)}
\gpcolor{color=gpbgfillcolor}
\gp3point{gp mark 7}{}{(1.550,1.198)}
\gpcolor{rgb color={0.863,0.149,0.498}}
\gp3point{gp mark 7}{}{(1.550,1.198)}
\gpcolor{color=gpbgfillcolor}
\gp3point{gp mark 7}{}{(1.797,1.387)}
\gpcolor{rgb color={0.863,0.149,0.498}}
\gp3point{gp mark 7}{}{(1.797,1.387)}
\gpcolor{color=gpbgfillcolor}
\gp3point{gp mark 7}{}{(2.044,1.575)}
\gpcolor{rgb color={0.863,0.149,0.498}}
\gp3point{gp mark 7}{}{(2.044,1.575)}
\gpcolor{color=gpbgfillcolor}
\gp3point{gp mark 7}{}{(2.290,1.752)}
\gpcolor{rgb color={0.863,0.149,0.498}}
\gp3point{gp mark 7}{}{(2.290,1.752)}
\gpcolor{color=gpbgfillcolor}
\gp3point{gp mark 7}{}{(2.537,1.901)}
\gpcolor{rgb color={0.863,0.149,0.498}}
\gp3point{gp mark 7}{}{(2.537,1.901)}
\gpcolor{color=gpbgfillcolor}
\gp3point{gp mark 7}{}{(2.784,2.015)}
\gpcolor{rgb color={0.863,0.149,0.498}}
\gp3point{gp mark 7}{}{(2.784,2.015)}
\gpcolor{color=gpbgfillcolor}
\gp3point{gp mark 7}{}{(3.030,2.048)}
\gpcolor{rgb color={0.863,0.149,0.498}}
\gp3point{gp mark 7}{}{(3.030,2.048)}
\gpcolor{color=gpbgfillcolor}
\gp3point{gp mark 7}{}{(3.277,2.200)}
\gpcolor{rgb color={0.863,0.149,0.498}}
\gp3point{gp mark 7}{}{(3.277,2.200)}
\gpcolor{color=gpbgfillcolor}
\gp3point{gp mark 7}{}{(3.524,2.350)}
\gpcolor{rgb color={0.863,0.149,0.498}}
\gp3point{gp mark 7}{}{(3.524,2.350)}
\gpcolor{color=gpbgfillcolor}
\gp3point{gp mark 7}{}{(3.770,2.492)}
\gpcolor{rgb color={0.863,0.149,0.498}}
\gp3point{gp mark 7}{}{(3.770,2.492)}
\gpsetlinewidth{2.00}
\draw[gp path] (1.304,1.023)--(1.550,1.198)--(1.797,1.387)--(2.044,1.575)--(2.290,1.752)%
  --(2.537,1.901)--(2.784,2.015)--(3.030,2.048)--(3.277,2.200)--(3.524,2.350)--(3.770,2.492);
\gp3point{gp mark 7}{}{(1.304,1.023)}
\gp3point{gp mark 7}{}{(1.550,1.198)}
\gp3point{gp mark 7}{}{(1.797,1.387)}
\gp3point{gp mark 7}{}{(2.044,1.575)}
\gp3point{gp mark 7}{}{(2.290,1.752)}
\gp3point{gp mark 7}{}{(2.537,1.901)}
\gp3point{gp mark 7}{}{(2.784,2.015)}
\gp3point{gp mark 7}{}{(3.030,2.048)}
\gp3point{gp mark 7}{}{(3.277,2.200)}
\gp3point{gp mark 7}{}{(3.524,2.350)}
\gp3point{gp mark 7}{}{(3.770,2.492)}
\gpcolor{color=gp lt color border}
\draw[gp path] (1.057,2.667)--(1.057,0.675)--(4.017,0.675)--(4.017,2.667)--cycle;
\node[gp node center,rotate=-270.0] at (0.235,1.671) {Tput [Mops/s]};
\node[gp node center] at (2.537,0.060) {Threads};
\node[gp node center] at (2.537,2.870) {Scalar};
\node[gp node center] at (6.681,2.870) {AVX};
\gpdefrectangularnode{gp plot 1}{\pgfpoint{1.057cm}{0.675cm}}{\pgfpoint{4.017cm}{2.667cm}}
\gpcolor{color=gp lt color axes}
\gpsetlinetype{gp lt axes}
\gpsetdashtype{gp dt axes}
\gpsetlinewidth{0.80}
\draw[gp path] (5.201,0.675)--(8.118,0.675);
\gpcolor{color=gp lt color border}
\gpsetlinetype{gp lt border}
\gpsetdashtype{gp dt solid}
\gpsetlinewidth{2.00}
\draw[gp path] (5.201,0.675)--(5.291,0.675);
\draw[gp path] (8.118,0.675)--(8.028,0.675);
\node[gp node right] at (5.127,0.675) {$0$};
\gpcolor{color=gp lt color axes}
\gpsetlinetype{gp lt axes}
\gpsetdashtype{gp dt axes}
\gpsetlinewidth{0.80}
\draw[gp path] (5.201,1.339)--(8.118,1.339);
\gpcolor{color=gp lt color border}
\gpsetlinetype{gp lt border}
\gpsetdashtype{gp dt solid}
\gpsetlinewidth{2.00}
\draw[gp path] (5.201,1.339)--(5.291,1.339);
\draw[gp path] (8.118,1.339)--(8.028,1.339);
\node[gp node right] at (5.127,1.339) {$500$};
\gpcolor{color=gp lt color axes}
\gpsetlinetype{gp lt axes}
\gpsetdashtype{gp dt axes}
\gpsetlinewidth{0.80}
\draw[gp path] (5.201,2.003)--(8.118,2.003);
\gpcolor{color=gp lt color border}
\gpsetlinetype{gp lt border}
\gpsetdashtype{gp dt solid}
\gpsetlinewidth{2.00}
\draw[gp path] (5.201,2.003)--(5.291,2.003);
\draw[gp path] (8.118,2.003)--(8.028,2.003);
\node[gp node right] at (5.127,2.003) {$1000$};
\gpcolor{color=gp lt color axes}
\gpsetlinetype{gp lt axes}
\gpsetdashtype{gp dt axes}
\gpsetlinewidth{0.80}
\draw[gp path] (5.201,2.667)--(8.118,2.667);
\gpcolor{color=gp lt color border}
\gpsetlinetype{gp lt border}
\gpsetdashtype{gp dt solid}
\gpsetlinewidth{2.00}
\draw[gp path] (5.201,2.667)--(5.291,2.667);
\draw[gp path] (8.118,2.667)--(8.028,2.667);
\node[gp node right] at (5.127,2.667) {$1500$};
\gpcolor{color=gp lt color axes}
\gpsetlinetype{gp lt axes}
\gpsetdashtype{gp dt axes}
\gpsetlinewidth{0.80}
\draw[gp path] (5.201,0.675)--(5.201,2.667);
\gpcolor{color=gp lt color border}
\gpsetlinetype{gp lt border}
\gpsetdashtype{gp dt solid}
\gpsetlinewidth{2.00}
\draw[gp path] (5.201,0.675)--(5.201,0.765);
\node[gp node center] at (5.201,0.429) {$0$};
\gpcolor{color=gp lt color axes}
\gpsetlinetype{gp lt axes}
\gpsetdashtype{gp dt axes}
\gpsetlinewidth{0.80}
\draw[gp path] (6.173,0.675)--(6.173,2.667);
\gpcolor{color=gp lt color border}
\gpsetlinetype{gp lt border}
\gpsetdashtype{gp dt solid}
\gpsetlinewidth{2.00}
\draw[gp path] (6.173,0.675)--(6.173,0.765);
\node[gp node center] at (6.173,0.429) {$32$};
\gpcolor{color=gp lt color axes}
\gpsetlinetype{gp lt axes}
\gpsetdashtype{gp dt axes}
\gpsetlinewidth{0.80}
\draw[gp path] (7.146,0.675)--(7.146,2.667);
\gpcolor{color=gp lt color border}
\gpsetlinetype{gp lt border}
\gpsetdashtype{gp dt solid}
\gpsetlinewidth{2.00}
\draw[gp path] (7.146,0.675)--(7.146,0.765);
\node[gp node center] at (7.146,0.429) {$64$};
\gpcolor{color=gp lt color axes}
\gpsetlinetype{gp lt axes}
\gpsetdashtype{gp dt axes}
\gpsetlinewidth{0.80}
\draw[gp path] (8.118,0.675)--(8.118,2.667);
\gpcolor{color=gp lt color border}
\gpsetlinetype{gp lt border}
\gpsetdashtype{gp dt solid}
\gpsetlinewidth{2.00}
\draw[gp path] (8.118,0.675)--(8.118,0.765);
\node[gp node center] at (8.118,0.429) {$96$};
\draw[gp path] (5.201,2.667)--(5.201,0.675)--(8.118,0.675)--(8.118,2.667)--cycle;
\gpcolor{color=gpbgfillcolor}
\gpsetlinewidth{1.00}
\gp3point{gp mark 7}{}{(5.444,1.060)}
\gpcolor{rgb color={0.863,0.149,0.498}}
\gp3point{gp mark 7}{}{(5.444,1.060)}
\gpcolor{color=gpbgfillcolor}
\gp3point{gp mark 7}{}{(5.687,1.235)}
\gpcolor{rgb color={0.863,0.149,0.498}}
\gp3point{gp mark 7}{}{(5.687,1.235)}
\gpcolor{color=gpbgfillcolor}
\gp3point{gp mark 7}{}{(5.930,1.435)}
\gpcolor{rgb color={0.863,0.149,0.498}}
\gp3point{gp mark 7}{}{(5.930,1.435)}
\gpcolor{color=gpbgfillcolor}
\gp3point{gp mark 7}{}{(6.173,1.636)}
\gpcolor{rgb color={0.863,0.149,0.498}}
\gp3point{gp mark 7}{}{(6.173,1.636)}
\gpcolor{color=gpbgfillcolor}
\gp3point{gp mark 7}{}{(6.416,1.826)}
\gpcolor{rgb color={0.863,0.149,0.498}}
\gp3point{gp mark 7}{}{(6.416,1.826)}
\gpcolor{color=gpbgfillcolor}
\gp3point{gp mark 7}{}{(6.660,1.970)}
\gpcolor{rgb color={0.863,0.149,0.498}}
\gp3point{gp mark 7}{}{(6.660,1.970)}
\gpcolor{color=gpbgfillcolor}
\gp3point{gp mark 7}{}{(6.903,2.000)}
\gpcolor{rgb color={0.863,0.149,0.498}}
\gp3point{gp mark 7}{}{(6.903,2.000)}
\gpcolor{color=gpbgfillcolor}
\gp3point{gp mark 7}{}{(7.146,2.005)}
\gpcolor{rgb color={0.863,0.149,0.498}}
\gp3point{gp mark 7}{}{(7.146,2.005)}
\gpcolor{color=gpbgfillcolor}
\gp3point{gp mark 7}{}{(7.389,2.146)}
\gpcolor{rgb color={0.863,0.149,0.498}}
\gp3point{gp mark 7}{}{(7.389,2.146)}
\gpcolor{color=gpbgfillcolor}
\gp3point{gp mark 7}{}{(7.632,2.295)}
\gpcolor{rgb color={0.863,0.149,0.498}}
\gp3point{gp mark 7}{}{(7.632,2.295)}
\gpcolor{color=gpbgfillcolor}
\gp3point{gp mark 7}{}{(7.875,2.407)}
\gpcolor{rgb color={0.863,0.149,0.498}}
\gp3point{gp mark 7}{}{(7.875,2.407)}
\gpsetlinewidth{2.00}
\draw[gp path] (5.444,1.060)--(5.687,1.235)--(5.930,1.435)--(6.173,1.636)--(6.416,1.826)%
  --(6.660,1.970)--(6.903,2.000)--(7.146,2.005)--(7.389,2.146)--(7.632,2.295)--(7.875,2.407);
\gp3point{gp mark 7}{}{(5.444,1.060)}
\gp3point{gp mark 7}{}{(5.687,1.235)}
\gp3point{gp mark 7}{}{(5.930,1.435)}
\gp3point{gp mark 7}{}{(6.173,1.636)}
\gp3point{gp mark 7}{}{(6.416,1.826)}
\gp3point{gp mark 7}{}{(6.660,1.970)}
\gp3point{gp mark 7}{}{(6.903,2.000)}
\gp3point{gp mark 7}{}{(7.146,2.005)}
\gp3point{gp mark 7}{}{(7.389,2.146)}
\gp3point{gp mark 7}{}{(7.632,2.295)}
\gp3point{gp mark 7}{}{(7.875,2.407)}
\gpcolor{color=gp lt color border}
\draw[gp path] (5.201,2.667)--(5.201,0.675)--(8.118,0.675)--(8.118,2.667)--cycle;
\node[gp node center,rotate=-270.0] at (4.232,1.671) {Tput [Mops/s]};
\node[gp node center] at (6.659,0.060) {Threads};
\node[gp node center] at (2.537,2.870) {Scalar};
\node[gp node center] at (6.681,2.870) {AVX};
\gpdefrectangularnode{gp plot 2}{\pgfpoint{5.201cm}{0.675cm}}{\pgfpoint{8.118cm}{2.667cm}}
\end{tikzpicture}
\caption{Aggregate compute throughput as the workload increases
  the number of cores. Throughput briefly flattens near the package's power
  and thermal knee, after which each additional core contributes less work.}%
\Description{Same as caption.}%
\label{fig:motiv:sweep-total}%
\end{figure}

\subsection{The Power and Thermal Knee}

Core count is a poor proxy for compute capacity once a processor approaches
its package limits. To expose this effect, we run a matrix
multiplication workload on an increasing number of physical cores.
\autoref{fig:motiv:sweep-total} reports the throughput for scalar and AVX baselines.
Initially, each additional core contributes substantial useful work. The curve
then reaches a knee in which throughput is nearly flat
while more cores are activated, before increasing again at a
lower rate.

\textbf{\textit{Package-level power and thermal limits reduce per-core performance.}}
Activating another core increases potential parallelism, but also forces the
processor to divide finite electrical and thermal headroom.
Dynamic voltage and frequency scaling responds by reducing core frequency.
Around the knee, where the processor starts to more aggressively downclock,
the loss on already-active cores can nearly cancel the work
provided by a newly activated core. Beyond it, throughput still grows, but the
incremental contribution of each core is smaller.
Before the knee, each additional core contributes an average of 4.24 million 
operations per second for scalar matrix multiplication. 
Beyond the knee plateau, that falls to 2.79 million operations per second. 
For AVX, the contribution drops from 22.6 to 12.6 million operations 
per second per additional core.

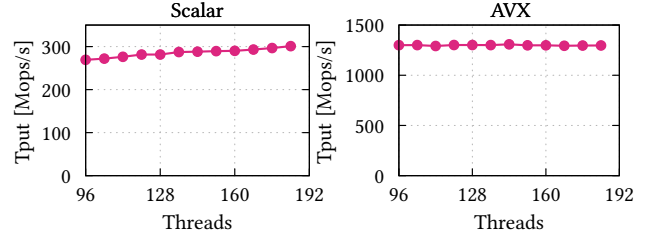
\begin{figure}%
\centering%
\begin{tikzpicture}[gnuplot]
\tikzset{every node/.append style={font={\fontsize{8.0pt}{9.6pt}\selectfont}}}
\path (0.000,0.000) rectangle (8.458,3.378);
\gpcolor{color=gp lt color axes}
\gpsetlinetype{gp lt axes}
\gpsetdashtype{gp dt axes}
\gpsetlinewidth{0.80}
\draw[gp path] (1.057,0.675)--(4.017,0.675);
\gpcolor{color=gp lt color border}
\gpsetlinetype{gp lt border}
\gpsetdashtype{gp dt solid}
\gpsetlinewidth{2.00}
\draw[gp path] (1.057,0.675)--(1.147,0.675);
\draw[gp path] (4.017,0.675)--(3.927,0.675);
\node[gp node right] at (0.983,0.675) {$0$};
\gpcolor{color=gp lt color axes}
\gpsetlinetype{gp lt axes}
\gpsetdashtype{gp dt axes}
\gpsetlinewidth{0.80}
\draw[gp path] (1.057,1.244)--(4.017,1.244);
\gpcolor{color=gp lt color border}
\gpsetlinetype{gp lt border}
\gpsetdashtype{gp dt solid}
\gpsetlinewidth{2.00}
\draw[gp path] (1.057,1.244)--(1.147,1.244);
\draw[gp path] (4.017,1.244)--(3.927,1.244);
\node[gp node right] at (0.983,1.244) {$100$};
\gpcolor{color=gp lt color axes}
\gpsetlinetype{gp lt axes}
\gpsetdashtype{gp dt axes}
\gpsetlinewidth{0.80}
\draw[gp path] (1.057,1.813)--(4.017,1.813);
\gpcolor{color=gp lt color border}
\gpsetlinetype{gp lt border}
\gpsetdashtype{gp dt solid}
\gpsetlinewidth{2.00}
\draw[gp path] (1.057,1.813)--(1.147,1.813);
\draw[gp path] (4.017,1.813)--(3.927,1.813);
\node[gp node right] at (0.983,1.813) {$200$};
\gpcolor{color=gp lt color axes}
\gpsetlinetype{gp lt axes}
\gpsetdashtype{gp dt axes}
\gpsetlinewidth{0.80}
\draw[gp path] (1.057,2.382)--(4.017,2.382);
\gpcolor{color=gp lt color border}
\gpsetlinetype{gp lt border}
\gpsetdashtype{gp dt solid}
\gpsetlinewidth{2.00}
\draw[gp path] (1.057,2.382)--(1.147,2.382);
\draw[gp path] (4.017,2.382)--(3.927,2.382);
\node[gp node right] at (0.983,2.382) {$300$};
\gpcolor{color=gp lt color axes}
\gpsetlinetype{gp lt axes}
\gpsetdashtype{gp dt axes}
\gpsetlinewidth{0.80}
\draw[gp path] (1.057,0.675)--(1.057,2.667);
\gpcolor{color=gp lt color border}
\gpsetlinetype{gp lt border}
\gpsetdashtype{gp dt solid}
\gpsetlinewidth{2.00}
\draw[gp path] (1.057,0.675)--(1.057,0.765);
\node[gp node center] at (1.057,0.429) {$96$};
\gpcolor{color=gp lt color axes}
\gpsetlinetype{gp lt axes}
\gpsetdashtype{gp dt axes}
\gpsetlinewidth{0.80}
\draw[gp path] (2.044,0.675)--(2.044,2.667);
\gpcolor{color=gp lt color border}
\gpsetlinetype{gp lt border}
\gpsetdashtype{gp dt solid}
\gpsetlinewidth{2.00}
\draw[gp path] (2.044,0.675)--(2.044,0.765);
\node[gp node center] at (2.044,0.429) {$128$};
\gpcolor{color=gp lt color axes}
\gpsetlinetype{gp lt axes}
\gpsetdashtype{gp dt axes}
\gpsetlinewidth{0.80}
\draw[gp path] (3.030,0.675)--(3.030,2.667);
\gpcolor{color=gp lt color border}
\gpsetlinetype{gp lt border}
\gpsetdashtype{gp dt solid}
\gpsetlinewidth{2.00}
\draw[gp path] (3.030,0.675)--(3.030,0.765);
\node[gp node center] at (3.030,0.429) {$160$};
\gpcolor{color=gp lt color axes}
\gpsetlinetype{gp lt axes}
\gpsetdashtype{gp dt axes}
\gpsetlinewidth{0.80}
\draw[gp path] (4.017,0.675)--(4.017,2.667);
\gpcolor{color=gp lt color border}
\gpsetlinetype{gp lt border}
\gpsetdashtype{gp dt solid}
\gpsetlinewidth{2.00}
\draw[gp path] (4.017,0.675)--(4.017,0.765);
\node[gp node center] at (4.017,0.429) {$192$};
\draw[gp path] (1.057,2.667)--(1.057,0.675)--(4.017,0.675)--(4.017,2.667)--cycle;
\gpcolor{rgb color={0.863,0.149,0.498}}
\gpsetlinewidth{1.00}
\draw[gp path] (1.057,2.177)--(1.057,2.237);
\draw[gp path] (1.304,2.208)--(1.304,2.237);
\draw[gp path] (1.550,2.246)--(1.550,2.248);
\draw[gp path] (1.797,2.271)--(1.797,2.282);
\draw[gp path] (2.044,2.274)--(2.044,2.280);
\draw[gp path] (2.290,2.307)--(2.290,2.312);
\draw[gp path] (2.537,2.307)--(2.537,2.323);
\draw[gp path] (2.784,2.320)--(2.784,2.323);
\draw[gp path] (3.030,2.309)--(3.030,2.343);
\draw[gp path] (3.277,2.326)--(3.277,2.359);
\draw[gp path] (3.524,2.359)--(3.524,2.367);
\draw[gp path] (3.770,2.382)--(3.770,2.395);
\gpcolor{color=gpbgfillcolor}
\gpsetpointsize{4.00}
\gp3point{gp mark 7}{}{(1.057,2.207)}
\gpcolor{rgb color={0.863,0.149,0.498}}
\gp3point{gp mark 7}{}{(1.057,2.207)}
\gpcolor{color=gpbgfillcolor}
\gp3point{gp mark 7}{}{(1.304,2.223)}
\gpcolor{rgb color={0.863,0.149,0.498}}
\gp3point{gp mark 7}{}{(1.304,2.223)}
\gpcolor{color=gpbgfillcolor}
\gp3point{gp mark 7}{}{(1.550,2.247)}
\gpcolor{rgb color={0.863,0.149,0.498}}
\gp3point{gp mark 7}{}{(1.550,2.247)}
\gpcolor{color=gpbgfillcolor}
\gp3point{gp mark 7}{}{(1.797,2.277)}
\gpcolor{rgb color={0.863,0.149,0.498}}
\gp3point{gp mark 7}{}{(1.797,2.277)}
\gpcolor{color=gpbgfillcolor}
\gp3point{gp mark 7}{}{(2.044,2.277)}
\gpcolor{rgb color={0.863,0.149,0.498}}
\gp3point{gp mark 7}{}{(2.044,2.277)}
\gpcolor{color=gpbgfillcolor}
\gp3point{gp mark 7}{}{(2.290,2.310)}
\gpcolor{rgb color={0.863,0.149,0.498}}
\gp3point{gp mark 7}{}{(2.290,2.310)}
\gpcolor{color=gpbgfillcolor}
\gp3point{gp mark 7}{}{(2.537,2.315)}
\gpcolor{rgb color={0.863,0.149,0.498}}
\gp3point{gp mark 7}{}{(2.537,2.315)}
\gpcolor{color=gpbgfillcolor}
\gp3point{gp mark 7}{}{(2.784,2.322)}
\gpcolor{rgb color={0.863,0.149,0.498}}
\gp3point{gp mark 7}{}{(2.784,2.322)}
\gpcolor{color=gpbgfillcolor}
\gp3point{gp mark 7}{}{(3.030,2.326)}
\gpcolor{rgb color={0.863,0.149,0.498}}
\gp3point{gp mark 7}{}{(3.030,2.326)}
\gpcolor{color=gpbgfillcolor}
\gp3point{gp mark 7}{}{(3.277,2.343)}
\gpcolor{rgb color={0.863,0.149,0.498}}
\gp3point{gp mark 7}{}{(3.277,2.343)}
\gpcolor{color=gpbgfillcolor}
\gp3point{gp mark 7}{}{(3.524,2.363)}
\gpcolor{rgb color={0.863,0.149,0.498}}
\gp3point{gp mark 7}{}{(3.524,2.363)}
\gpcolor{color=gpbgfillcolor}
\gp3point{gp mark 7}{}{(3.770,2.388)}
\gpcolor{rgb color={0.863,0.149,0.498}}
\gp3point{gp mark 7}{}{(3.770,2.388)}
\gpsetlinewidth{2.00}
\draw[gp path] (1.057,2.207)--(1.304,2.223)--(1.550,2.247)--(1.797,2.277)--(2.044,2.277)%
  --(2.290,2.310)--(2.537,2.315)--(2.784,2.322)--(3.030,2.326)--(3.277,2.343)--(3.524,2.363)%
  --(3.770,2.388);
\gp3point{gp mark 7}{}{(1.057,2.207)}
\gp3point{gp mark 7}{}{(1.304,2.223)}
\gp3point{gp mark 7}{}{(1.550,2.247)}
\gp3point{gp mark 7}{}{(1.797,2.277)}
\gp3point{gp mark 7}{}{(2.044,2.277)}
\gp3point{gp mark 7}{}{(2.290,2.310)}
\gp3point{gp mark 7}{}{(2.537,2.315)}
\gp3point{gp mark 7}{}{(2.784,2.322)}
\gp3point{gp mark 7}{}{(3.030,2.326)}
\gp3point{gp mark 7}{}{(3.277,2.343)}
\gp3point{gp mark 7}{}{(3.524,2.363)}
\gp3point{gp mark 7}{}{(3.770,2.388)}
\gpcolor{color=gp lt color border}
\draw[gp path] (1.057,2.667)--(1.057,0.675)--(4.017,0.675)--(4.017,2.667)--cycle;
\node[gp node center,rotate=-270.0] at (0.235,1.671) {Tput [Mops/s]};
\node[gp node center] at (2.537,0.060) {Threads};
\node[gp node center] at (2.537,2.870) {Scalar};
\node[gp node center] at (6.681,2.870) {AVX};
\gpdefrectangularnode{gp plot 1}{\pgfpoint{1.057cm}{0.675cm}}{\pgfpoint{4.017cm}{2.667cm}}
\gpcolor{color=gp lt color axes}
\gpsetlinetype{gp lt axes}
\gpsetdashtype{gp dt axes}
\gpsetlinewidth{0.80}
\draw[gp path] (5.201,0.675)--(8.118,0.675);
\gpcolor{color=gp lt color border}
\gpsetlinetype{gp lt border}
\gpsetdashtype{gp dt solid}
\gpsetlinewidth{2.00}
\draw[gp path] (5.201,0.675)--(5.291,0.675);
\draw[gp path] (8.118,0.675)--(8.028,0.675);
\node[gp node right] at (5.127,0.675) {$0$};
\gpcolor{color=gp lt color axes}
\gpsetlinetype{gp lt axes}
\gpsetdashtype{gp dt axes}
\gpsetlinewidth{0.80}
\draw[gp path] (5.201,1.339)--(8.118,1.339);
\gpcolor{color=gp lt color border}
\gpsetlinetype{gp lt border}
\gpsetdashtype{gp dt solid}
\gpsetlinewidth{2.00}
\draw[gp path] (5.201,1.339)--(5.291,1.339);
\draw[gp path] (8.118,1.339)--(8.028,1.339);
\node[gp node right] at (5.127,1.339) {$500$};
\gpcolor{color=gp lt color axes}
\gpsetlinetype{gp lt axes}
\gpsetdashtype{gp dt axes}
\gpsetlinewidth{0.80}
\draw[gp path] (5.201,2.003)--(8.118,2.003);
\gpcolor{color=gp lt color border}
\gpsetlinetype{gp lt border}
\gpsetdashtype{gp dt solid}
\gpsetlinewidth{2.00}
\draw[gp path] (5.201,2.003)--(5.291,2.003);
\draw[gp path] (8.118,2.003)--(8.028,2.003);
\node[gp node right] at (5.127,2.003) {$1000$};
\gpcolor{color=gp lt color axes}
\gpsetlinetype{gp lt axes}
\gpsetdashtype{gp dt axes}
\gpsetlinewidth{0.80}
\draw[gp path] (5.201,2.667)--(8.118,2.667);
\gpcolor{color=gp lt color border}
\gpsetlinetype{gp lt border}
\gpsetdashtype{gp dt solid}
\gpsetlinewidth{2.00}
\draw[gp path] (5.201,2.667)--(5.291,2.667);
\draw[gp path] (8.118,2.667)--(8.028,2.667);
\node[gp node right] at (5.127,2.667) {$1500$};
\gpcolor{color=gp lt color axes}
\gpsetlinetype{gp lt axes}
\gpsetdashtype{gp dt axes}
\gpsetlinewidth{0.80}
\draw[gp path] (5.201,0.675)--(5.201,2.667);
\gpcolor{color=gp lt color border}
\gpsetlinetype{gp lt border}
\gpsetdashtype{gp dt solid}
\gpsetlinewidth{2.00}
\draw[gp path] (5.201,0.675)--(5.201,0.765);
\node[gp node center] at (5.201,0.429) {$96$};
\gpcolor{color=gp lt color axes}
\gpsetlinetype{gp lt axes}
\gpsetdashtype{gp dt axes}
\gpsetlinewidth{0.80}
\draw[gp path] (6.173,0.675)--(6.173,2.667);
\gpcolor{color=gp lt color border}
\gpsetlinetype{gp lt border}
\gpsetdashtype{gp dt solid}
\gpsetlinewidth{2.00}
\draw[gp path] (6.173,0.675)--(6.173,0.765);
\node[gp node center] at (6.173,0.429) {$128$};
\gpcolor{color=gp lt color axes}
\gpsetlinetype{gp lt axes}
\gpsetdashtype{gp dt axes}
\gpsetlinewidth{0.80}
\draw[gp path] (7.146,0.675)--(7.146,2.667);
\gpcolor{color=gp lt color border}
\gpsetlinetype{gp lt border}
\gpsetdashtype{gp dt solid}
\gpsetlinewidth{2.00}
\draw[gp path] (7.146,0.675)--(7.146,0.765);
\node[gp node center] at (7.146,0.429) {$160$};
\gpcolor{color=gp lt color axes}
\gpsetlinetype{gp lt axes}
\gpsetdashtype{gp dt axes}
\gpsetlinewidth{0.80}
\draw[gp path] (8.118,0.675)--(8.118,2.667);
\gpcolor{color=gp lt color border}
\gpsetlinetype{gp lt border}
\gpsetdashtype{gp dt solid}
\gpsetlinewidth{2.00}
\draw[gp path] (8.118,0.675)--(8.118,0.765);
\node[gp node center] at (8.118,0.429) {$192$};
\draw[gp path] (5.201,2.667)--(5.201,0.675)--(8.118,0.675)--(8.118,2.667)--cycle;
\gpcolor{rgb color={0.863,0.149,0.498}}
\gpsetlinewidth{1.00}
\draw[gp path] (6.416,2.402)--(6.416,2.403);
\gpcolor{color=gpbgfillcolor}
\gp3point{gp mark 7}{}{(5.201,2.401)}
\gpcolor{rgb color={0.863,0.149,0.498}}
\gp3point{gp mark 7}{}{(5.201,2.401)}
\gpcolor{color=gpbgfillcolor}
\gp3point{gp mark 7}{}{(5.444,2.402)}
\gpcolor{rgb color={0.863,0.149,0.498}}
\gp3point{gp mark 7}{}{(5.444,2.402)}
\gpcolor{color=gpbgfillcolor}
\gp3point{gp mark 7}{}{(5.687,2.390)}
\gpcolor{rgb color={0.863,0.149,0.498}}
\gp3point{gp mark 7}{}{(5.687,2.390)}
\gpcolor{color=gpbgfillcolor}
\gp3point{gp mark 7}{}{(5.930,2.403)}
\gpcolor{rgb color={0.863,0.149,0.498}}
\gp3point{gp mark 7}{}{(5.930,2.403)}
\gpcolor{color=gpbgfillcolor}
\gp3point{gp mark 7}{}{(6.173,2.403)}
\gpcolor{rgb color={0.863,0.149,0.498}}
\gp3point{gp mark 7}{}{(6.173,2.403)}
\gpcolor{color=gpbgfillcolor}
\gp3point{gp mark 7}{}{(6.416,2.402)}
\gpcolor{rgb color={0.863,0.149,0.498}}
\gp3point{gp mark 7}{}{(6.416,2.402)}
\gpcolor{color=gpbgfillcolor}
\gp3point{gp mark 7}{}{(6.660,2.412)}
\gpcolor{rgb color={0.863,0.149,0.498}}
\gp3point{gp mark 7}{}{(6.660,2.412)}
\gpcolor{color=gpbgfillcolor}
\gp3point{gp mark 7}{}{(6.903,2.399)}
\gpcolor{rgb color={0.863,0.149,0.498}}
\gp3point{gp mark 7}{}{(6.903,2.399)}
\gpcolor{color=gpbgfillcolor}
\gp3point{gp mark 7}{}{(7.146,2.399)}
\gpcolor{rgb color={0.863,0.149,0.498}}
\gp3point{gp mark 7}{}{(7.146,2.399)}
\gpcolor{color=gpbgfillcolor}
\gp3point{gp mark 7}{}{(7.389,2.393)}
\gpcolor{rgb color={0.863,0.149,0.498}}
\gp3point{gp mark 7}{}{(7.389,2.393)}
\gpcolor{color=gpbgfillcolor}
\gp3point{gp mark 7}{}{(7.632,2.396)}
\gpcolor{rgb color={0.863,0.149,0.498}}
\gp3point{gp mark 7}{}{(7.632,2.396)}
\gpcolor{color=gpbgfillcolor}
\gp3point{gp mark 7}{}{(7.875,2.397)}
\gpcolor{rgb color={0.863,0.149,0.498}}
\gp3point{gp mark 7}{}{(7.875,2.397)}
\gpsetlinewidth{2.00}
\draw[gp path] (5.201,2.401)--(5.444,2.402)--(5.687,2.390)--(5.930,2.403)--(6.173,2.403)%
  --(6.416,2.402)--(6.660,2.412)--(6.903,2.399)--(7.146,2.399)--(7.389,2.393)--(7.632,2.396)%
  --(7.875,2.397);
\gp3point{gp mark 7}{}{(5.201,2.401)}
\gp3point{gp mark 7}{}{(5.444,2.402)}
\gp3point{gp mark 7}{}{(5.687,2.390)}
\gp3point{gp mark 7}{}{(5.930,2.403)}
\gp3point{gp mark 7}{}{(6.173,2.403)}
\gp3point{gp mark 7}{}{(6.416,2.402)}
\gp3point{gp mark 7}{}{(6.660,2.412)}
\gp3point{gp mark 7}{}{(6.903,2.399)}
\gp3point{gp mark 7}{}{(7.146,2.399)}
\gp3point{gp mark 7}{}{(7.389,2.393)}
\gp3point{gp mark 7}{}{(7.632,2.396)}
\gp3point{gp mark 7}{}{(7.875,2.397)}
\gpcolor{color=gp lt color border}
\draw[gp path] (5.201,2.667)--(5.201,0.675)--(8.118,0.675)--(8.118,2.667)--cycle;
\node[gp node center,rotate=-270.0] at (4.232,1.671) {Tput [Mops/s]};
\node[gp node center] at (6.659,0.060) {Threads};
\node[gp node center] at (2.537,2.870) {Scalar};
\node[gp node center] at (6.681,2.870) {AVX};
\gpdefrectangularnode{gp plot 2}{\pgfpoint{5.201cm}{0.675cm}}{\pgfpoint{8.118cm}{2.667cm}}
\end{tikzpicture}
\caption{Aggregate compute throughput flattens once 
  the workload scales beyond physical cores to SMT.}%
\Description{Same as caption.}%
\label{fig:motiv:sweep-total-smt}%
\end{figure}

\textbf{\textit{SMT threads provide little additional compute.}}
\autoref{fig:motiv:sweep-total-smt} shows that once the processor exhausts
its physical cores, SMT threads contribute even less additional compute.
For compute-intensive workloads, such as AVX, aggregate throughput
nearly plateaus despite activating more hardware threads, while scalar workloads
see minimal throughput increase.
Here adding cores provides diminishing returns because they compete
for the same package budget.

\par\medskip
\noindent\fbox{%
\parbox{\dimexpr\linewidth-2\fboxsep-2\fboxrule\relax}{%
\normalsize
\textbf{Fixed-Budget Model.} Below the processor package limit,
additional active cores increase compute.
Above this limit, additional cores yield marginal performance gains.
Activity on one core reduces the budget available to others.
Idling offers that budget back for redistribution.
}}
\par\smallskip

\subsection{The Waiting Spectrum}

\begin{figure}
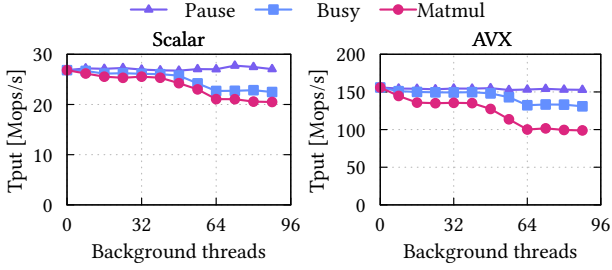
%
\centering%

\caption{Foreground matrix multiplication throughput as increasing core counts
  perform background 
  work (\texttt{Matmul}, \texttt{Busy}, or \texttt{Pause}).
  \texttt{Pause} preserves foreground
  capacity, whereas compute and busy polling consume the shared package budget.}%
\Description{Same as caption.}%
\label{fig:motiv:sweep-foreground}%
\end{figure}

\begin{figure}
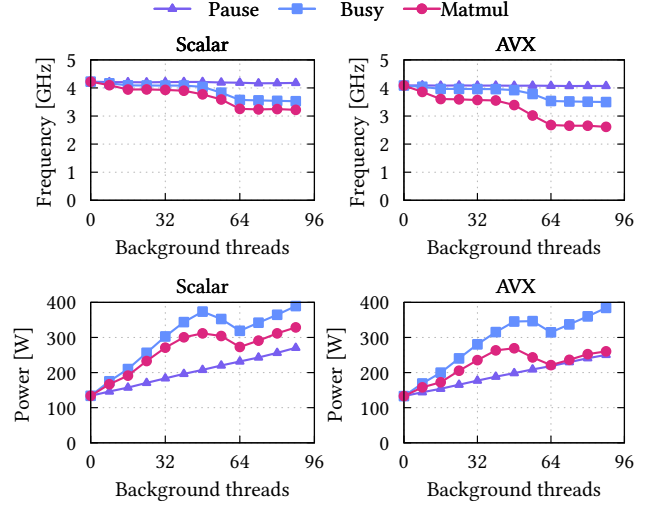
%
\centering%
%
\caption{Foreground core frequency (top) and package power (bottom)
  as cores are assigned to background 
  work (\texttt{Matmul}, \texttt{Busy}, or \texttt{Pause}).
  \texttt{Pause} polling preserves foreground frequency and uses less package
  power than busy polling.}%
\Description{Same as caption.}%
\label{fig:motiv:sweep-freq}%
\end{figure}

In the fixed-budget regime, a thread that enters hardware idle states
returns compute budget that the processor can redistribute to productive work.
We isolate this redistribution with an eight-thread matrix multiplication
foreground workload. We then place an increasing number of background threads
on otherwise unused physical cores. The background either performs matrix
multiplication or polls queues, akin to a low-latency network stack,
with different waiting strategies.
\autoref{fig:motiv:sweep-foreground} reports foreground throughput, while
\autoref{fig:motiv:sweep-freq} reports core frequency and total power.

\textbf{\textit{Waiting strategy determines foreground application performance.}}
Scalar compute-heavy background work
reduces foreground throughput from 26.8 to 20.5~Mops/s, and busy polling
reduces it to 22.5~Mops/s. In both cases, package activity consumes headroom and the
processor downclocks the foreground. By contrast, \texttt{Pause}
polling retains foreground throughput and frequency while using
much less power than \texttt{Busy}, leaving headroom for the foreground.

\textbf{\textit{AVX workloads magnify the cost of budget contention.}}
AVX instructions amplify the effects of the fixed budget
as the chip reaches its limits. Throughput drops from 155.9 to 130.9~Mops/s
with background busy polling and falls further to 98.9~Mops/s with
vectorised matrix multiplication. \texttt{Pause} polling
again preserves foreground throughput with the same eight-core allocation
as more cores are added.

\par\medskip
\noindent\fbox{%
\parbox{\dimexpr\linewidth-2\fboxsep-2\fboxrule\relax}{%
\normalsize
\textbf{L1: Idleness transfers power and thermal headroom.}
An idle service core does not 
necessarily waste compute. Efficient
waiting returns power and thermal headroom that hardware
redistributes to productive work.
}}
\par\smallskip

\subsection{Placement Determines Budget Sharing}

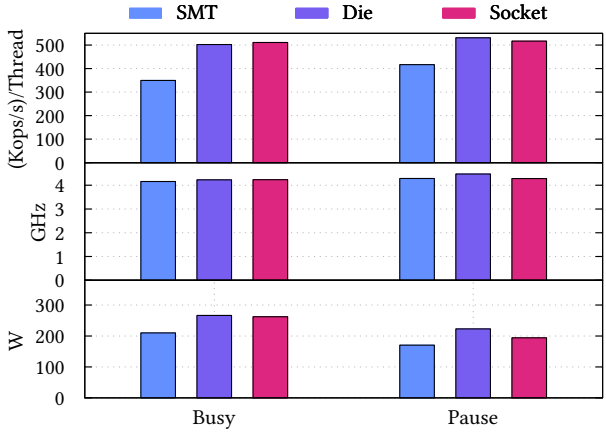
\begin{figure}%
\centering%
\begin{tikzpicture}[gnuplot]
\tikzset{every node/.append style={font={\fontsize{8.0pt}{9.6pt}\selectfont}}}
\path (0.000,0.000) rectangle (8.458,2.794);
\gpcolor{color=gp lt color axes}
\gpsetlinetype{gp lt axes}
\gpsetdashtype{gp dt axes}
\gpsetlinewidth{0.40}
\draw[gp path] (1.310,0.572)--(2.917,0.572);
\gpcolor{color=gp lt color border}
\gpsetlinetype{gp lt border}
\gpsetdashtype{gp dt solid}
\gpsetlinewidth{2.00}
\draw[gp path] (1.310,0.572)--(1.400,0.572);
\draw[gp path] (2.917,0.572)--(2.827,0.572);
\node[gp node right] at (1.236,0.572) {$0$};
\gpcolor{color=gp lt color axes}
\gpsetlinetype{gp lt axes}
\gpsetdashtype{gp dt axes}
\gpsetlinewidth{0.40}
\draw[gp path] (1.310,0.892)--(2.917,0.892);
\gpcolor{color=gp lt color border}
\gpsetlinetype{gp lt border}
\gpsetdashtype{gp dt solid}
\gpsetlinewidth{2.00}
\draw[gp path] (1.310,0.892)--(1.400,0.892);
\draw[gp path] (2.917,0.892)--(2.827,0.892);
\node[gp node right] at (1.236,0.892) {$100$};
\gpcolor{color=gp lt color axes}
\gpsetlinetype{gp lt axes}
\gpsetdashtype{gp dt axes}
\gpsetlinewidth{0.40}
\draw[gp path] (1.310,1.212)--(2.917,1.212);
\gpcolor{color=gp lt color border}
\gpsetlinetype{gp lt border}
\gpsetdashtype{gp dt solid}
\gpsetlinewidth{2.00}
\draw[gp path] (1.310,1.212)--(1.400,1.212);
\draw[gp path] (2.917,1.212)--(2.827,1.212);
\node[gp node right] at (1.236,1.212) {$200$};
\gpcolor{color=gp lt color axes}
\gpsetlinetype{gp lt axes}
\gpsetdashtype{gp dt axes}
\gpsetlinewidth{0.40}
\draw[gp path] (1.310,1.532)--(2.917,1.532);
\gpcolor{color=gp lt color border}
\gpsetlinetype{gp lt border}
\gpsetdashtype{gp dt solid}
\gpsetlinewidth{2.00}
\draw[gp path] (1.310,1.532)--(1.400,1.532);
\draw[gp path] (2.917,1.532)--(2.827,1.532);
\node[gp node right] at (1.236,1.532) {$300$};
\gpcolor{color=gp lt color axes}
\gpsetlinetype{gp lt axes}
\gpsetdashtype{gp dt axes}
\gpsetlinewidth{0.40}
\draw[gp path] (1.310,1.852)--(2.917,1.852);
\gpcolor{color=gp lt color border}
\gpsetlinetype{gp lt border}
\gpsetdashtype{gp dt solid}
\gpsetlinewidth{2.00}
\draw[gp path] (1.310,1.852)--(1.400,1.852);
\draw[gp path] (2.917,1.852)--(2.827,1.852);
\node[gp node right] at (1.236,1.852) {$400$};
\gpcolor{color=gp lt color axes}
\gpsetlinetype{gp lt axes}
\gpsetdashtype{gp dt axes}
\gpsetlinewidth{0.40}
\draw[gp path] (1.310,2.172)--(2.917,2.172);
\gpcolor{color=gp lt color border}
\gpsetlinetype{gp lt border}
\gpsetdashtype{gp dt solid}
\gpsetlinewidth{2.00}
\draw[gp path] (1.310,2.172)--(1.400,2.172);
\draw[gp path] (2.917,2.172)--(2.827,2.172);
\node[gp node right] at (1.236,2.172) {$500$};
\gpcolor{color=gp lt color axes}
\gpsetlinetype{gp lt axes}
\gpsetdashtype{gp dt axes}
\gpsetlinewidth{0.40}
\draw[gp path] (1.712,0.572)--(1.712,2.332);
\gpcolor{color=gp lt color border}
\gpsetlinetype{gp lt border}
\gpsetdashtype{gp dt solid}
\gpsetlinewidth{2.00}
\draw[gp path] (1.712,0.572)--(1.712,0.662);
\node[gp node center] at (1.712,0.326) {Busy};
\gpcolor{color=gp lt color axes}
\gpsetlinetype{gp lt axes}
\gpsetdashtype{gp dt axes}
\gpsetlinewidth{0.40}
\draw[gp path] (2.515,0.572)--(2.515,2.332);
\gpcolor{color=gp lt color border}
\gpsetlinetype{gp lt border}
\gpsetdashtype{gp dt solid}
\gpsetlinewidth{2.00}
\draw[gp path] (2.515,0.572)--(2.515,0.662);
\node[gp node center] at (2.515,0.326) {Pause};
\draw[gp path] (1.310,2.332)--(1.310,0.572)--(2.917,0.572)--(2.917,2.332)--cycle;
\node[gp node left] at (2.326,2.597) {SMT};
\node[gp node left] at (4.524,2.597) {Die};
\node[gp node left] at (6.470,2.597) {Socket};
\gpfill{rgb color={0.392,0.561,1.000}} (1.455,0.572)--(1.568,0.572)--(1.568,1.692)--(1.455,1.692)--cycle;
\gpsetlinewidth{1.00}
\draw[gp path] (1.455,0.572)--(1.455,1.691)--(1.567,1.691)--(1.567,0.572)--cycle;
\gpfill{rgb color={0.392,0.561,1.000}} (2.258,0.572)--(2.372,0.572)--(2.372,1.907)--(2.258,1.907)--cycle;
\draw[gp path] (2.258,0.572)--(2.258,1.906)--(2.371,1.906)--(2.371,0.572)--cycle;
\gpfill{rgb color={0.471,0.369,0.941}} (1.656,0.572)--(1.769,0.572)--(1.769,2.181)--(1.656,2.181)--cycle;
\draw[gp path] (1.656,0.572)--(1.656,2.180)--(1.768,2.180)--(1.768,0.572)--cycle;
\gpfill{rgb color={0.471,0.369,0.941}} (2.459,0.572)--(2.572,0.572)--(2.572,2.272)--(2.459,2.272)--cycle;
\draw[gp path] (2.459,0.572)--(2.459,2.271)--(2.571,2.271)--(2.571,0.572)--cycle;
\gpfill{rgb color={0.863,0.149,0.498}} (1.856,0.572)--(1.970,0.572)--(1.970,2.208)--(1.856,2.208)--cycle;
\draw[gp path] (1.856,0.572)--(1.856,2.207)--(1.969,2.207)--(1.969,0.572)--cycle;
\gpfill{rgb color={0.863,0.149,0.498}} (2.660,0.572)--(2.773,0.572)--(2.773,2.227)--(2.660,2.227)--cycle;
\draw[gp path] (2.660,0.572)--(2.660,2.226)--(2.772,2.226)--(2.772,0.572)--cycle;
\gpsetlinewidth{2.00}
\draw[gp path] (1.310,2.332)--(1.310,0.572)--(2.917,0.572)--(2.917,2.332)--cycle;
\node[gp node center,rotate=-270.0] at (0.415,1.452) {(Kops/s)/Thread};
\gpfill{rgb color={0.392,0.561,1.000}} (1.818,2.547)--(2.240,2.547)--(2.240,2.633)--(1.818,2.633)--cycle;
\gpcolor{rgb color={0.392,0.561,1.000}}
\gpsetlinewidth{1.00}
\draw[gp path] (1.818,2.547)--(1.818,2.633)--(2.240,2.633)--(2.240,2.547)--cycle;
\gpfill{rgb color={0.471,0.369,0.941}} (4.017,2.547)--(4.439,2.547)--(4.439,2.633)--(4.017,2.633)--cycle;
\gpcolor{rgb color={0.471,0.369,0.941}}
\draw[gp path] (4.017,2.547)--(4.017,2.633)--(4.439,2.633)--(4.439,2.547)--cycle;
\gpfill{rgb color={0.863,0.149,0.498}} (5.962,2.547)--(6.384,2.547)--(6.384,2.633)--(5.962,2.633)--cycle;
\gpcolor{rgb color={0.863,0.149,0.498}}
\draw[gp path] (5.962,2.547)--(5.962,2.633)--(6.384,2.633)--(6.384,2.547)--cycle;
\gpdefrectangularnode{gp plot 1}{\pgfpoint{1.310cm}{0.572cm}}{\pgfpoint{2.917cm}{2.332cm}}
\gpcolor{color=gp lt color axes}
\gpsetlinetype{gp lt axes}
\gpsetdashtype{gp dt axes}
\gpsetlinewidth{0.40}
\draw[gp path] (4.017,0.572)--(5.623,0.572);
\gpcolor{color=gp lt color border}
\gpsetlinetype{gp lt border}
\gpsetdashtype{gp dt solid}
\gpsetlinewidth{2.00}
\draw[gp path] (4.017,0.572)--(4.107,0.572);
\draw[gp path] (5.623,0.572)--(5.533,0.572);
\node[gp node right] at (3.943,0.572) {$0$};
\gpcolor{color=gp lt color axes}
\gpsetlinetype{gp lt axes}
\gpsetdashtype{gp dt axes}
\gpsetlinewidth{0.40}
\draw[gp path] (4.017,0.928)--(5.623,0.928);
\gpcolor{color=gp lt color border}
\gpsetlinetype{gp lt border}
\gpsetdashtype{gp dt solid}
\gpsetlinewidth{2.00}
\draw[gp path] (4.017,0.928)--(4.107,0.928);
\draw[gp path] (5.623,0.928)--(5.533,0.928);
\node[gp node right] at (3.943,0.928) {$1$};
\gpcolor{color=gp lt color axes}
\gpsetlinetype{gp lt axes}
\gpsetdashtype{gp dt axes}
\gpsetlinewidth{0.40}
\draw[gp path] (4.017,1.283)--(5.623,1.283);
\gpcolor{color=gp lt color border}
\gpsetlinetype{gp lt border}
\gpsetdashtype{gp dt solid}
\gpsetlinewidth{2.00}
\draw[gp path] (4.017,1.283)--(4.107,1.283);
\draw[gp path] (5.623,1.283)--(5.533,1.283);
\node[gp node right] at (3.943,1.283) {$2$};
\gpcolor{color=gp lt color axes}
\gpsetlinetype{gp lt axes}
\gpsetdashtype{gp dt axes}
\gpsetlinewidth{0.40}
\draw[gp path] (4.017,1.639)--(5.623,1.639);
\gpcolor{color=gp lt color border}
\gpsetlinetype{gp lt border}
\gpsetdashtype{gp dt solid}
\gpsetlinewidth{2.00}
\draw[gp path] (4.017,1.639)--(4.107,1.639);
\draw[gp path] (5.623,1.639)--(5.533,1.639);
\node[gp node right] at (3.943,1.639) {$3$};
\gpcolor{color=gp lt color axes}
\gpsetlinetype{gp lt axes}
\gpsetdashtype{gp dt axes}
\gpsetlinewidth{0.40}
\draw[gp path] (4.017,1.994)--(5.623,1.994);
\gpcolor{color=gp lt color border}
\gpsetlinetype{gp lt border}
\gpsetdashtype{gp dt solid}
\gpsetlinewidth{2.00}
\draw[gp path] (4.017,1.994)--(4.107,1.994);
\draw[gp path] (5.623,1.994)--(5.533,1.994);
\node[gp node right] at (3.943,1.994) {$4$};
\gpcolor{color=gp lt color axes}
\gpsetlinetype{gp lt axes}
\gpsetdashtype{gp dt axes}
\gpsetlinewidth{0.40}
\draw[gp path] (4.419,0.572)--(4.419,2.332);
\gpcolor{color=gp lt color border}
\gpsetlinetype{gp lt border}
\gpsetdashtype{gp dt solid}
\gpsetlinewidth{2.00}
\draw[gp path] (4.419,0.572)--(4.419,0.662);
\node[gp node center] at (4.419,0.326) {Busy};
\gpcolor{color=gp lt color axes}
\gpsetlinetype{gp lt axes}
\gpsetdashtype{gp dt axes}
\gpsetlinewidth{0.40}
\draw[gp path] (5.222,0.572)--(5.222,2.332);
\gpcolor{color=gp lt color border}
\gpsetlinetype{gp lt border}
\gpsetdashtype{gp dt solid}
\gpsetlinewidth{2.00}
\draw[gp path] (5.222,0.572)--(5.222,0.662);
\node[gp node center] at (5.222,0.326) {Pause};
\draw[gp path] (4.017,2.332)--(4.017,0.572)--(5.623,0.572)--(5.623,2.332)--cycle;
\node[gp node left] at (2.326,2.597) {SMT};
\node[gp node left] at (4.524,2.597) {Die};
\node[gp node left] at (6.470,2.597) {Socket};
\gpfill{rgb color={0.392,0.561,1.000}} (4.162,0.572)--(4.275,0.572)--(4.275,2.051)--(4.162,2.051)--cycle;
\gpsetlinewidth{1.00}
\draw[gp path] (4.162,0.572)--(4.162,2.050)--(4.274,2.050)--(4.274,0.572)--cycle;
\gpfill{rgb color={0.392,0.561,1.000}} (4.965,0.572)--(5.078,0.572)--(5.078,2.096)--(4.965,2.096)--cycle;
\draw[gp path] (4.965,0.572)--(4.965,2.095)--(5.077,2.095)--(5.077,0.572)--cycle;
\gpfill{rgb color={0.471,0.369,0.941}} (4.362,0.572)--(4.476,0.572)--(4.476,2.076)--(4.362,2.076)--cycle;
\draw[gp path] (4.362,0.572)--(4.362,2.075)--(4.475,2.075)--(4.475,0.572)--cycle;
\gpfill{rgb color={0.471,0.369,0.941}} (5.165,0.572)--(5.279,0.572)--(5.279,2.163)--(5.165,2.163)--cycle;
\draw[gp path] (5.165,0.572)--(5.165,2.162)--(5.278,2.162)--(5.278,0.572)--cycle;
\gpfill{rgb color={0.863,0.149,0.498}} (4.563,0.572)--(4.676,0.572)--(4.676,2.078)--(4.563,2.078)--cycle;
\draw[gp path] (4.563,0.572)--(4.563,2.077)--(4.675,2.077)--(4.675,0.572)--cycle;
\gpfill{rgb color={0.863,0.149,0.498}} (5.366,0.572)--(5.479,0.572)--(5.479,2.095)--(5.366,2.095)--cycle;
\draw[gp path] (5.366,0.572)--(5.366,2.094)--(5.478,2.094)--(5.478,0.572)--cycle;
\gpsetlinewidth{2.00}
\draw[gp path] (4.017,2.332)--(4.017,0.572)--(5.623,0.572)--(5.623,2.332)--cycle;
\node[gp node center,rotate=-270.0] at (3.384,1.452) {GHz};
\gpfill{rgb color={0.392,0.561,1.000}} (1.818,2.547)--(2.240,2.547)--(2.240,2.633)--(1.818,2.633)--cycle;
\gpcolor{rgb color={0.392,0.561,1.000}}
\gpsetlinewidth{1.00}
\draw[gp path] (1.818,2.547)--(1.818,2.633)--(2.240,2.633)--(2.240,2.547)--cycle;
\gpfill{rgb color={0.471,0.369,0.941}} (4.017,2.547)--(4.439,2.547)--(4.439,2.633)--(4.017,2.633)--cycle;
\gpcolor{rgb color={0.471,0.369,0.941}}
\draw[gp path] (4.017,2.547)--(4.017,2.633)--(4.439,2.633)--(4.439,2.547)--cycle;
\gpfill{rgb color={0.863,0.149,0.498}} (5.962,2.547)--(6.384,2.547)--(6.384,2.633)--(5.962,2.633)--cycle;
\gpcolor{rgb color={0.863,0.149,0.498}}
\draw[gp path] (5.962,2.547)--(5.962,2.633)--(6.384,2.633)--(6.384,2.547)--cycle;
\gpdefrectangularnode{gp plot 2}{\pgfpoint{4.017cm}{0.572cm}}{\pgfpoint{5.623cm}{2.332cm}}
\gpcolor{color=gp lt color axes}
\gpsetlinetype{gp lt axes}
\gpsetdashtype{gp dt axes}
\gpsetlinewidth{0.40}
\draw[gp path] (6.724,0.572)--(8.330,0.572);
\gpcolor{color=gp lt color border}
\gpsetlinetype{gp lt border}
\gpsetdashtype{gp dt solid}
\gpsetlinewidth{2.00}
\draw[gp path] (6.724,0.572)--(6.814,0.572);
\draw[gp path] (8.330,0.572)--(8.240,0.572);
\node[gp node right] at (6.650,0.572) {$0$};
\gpcolor{color=gp lt color axes}
\gpsetlinetype{gp lt axes}
\gpsetdashtype{gp dt axes}
\gpsetlinewidth{0.40}
\draw[gp path] (6.724,1.035)--(8.330,1.035);
\gpcolor{color=gp lt color border}
\gpsetlinetype{gp lt border}
\gpsetdashtype{gp dt solid}
\gpsetlinewidth{2.00}
\draw[gp path] (6.724,1.035)--(6.814,1.035);
\draw[gp path] (8.330,1.035)--(8.240,1.035);
\node[gp node right] at (6.650,1.035) {$100$};
\gpcolor{color=gp lt color axes}
\gpsetlinetype{gp lt axes}
\gpsetdashtype{gp dt axes}
\gpsetlinewidth{0.40}
\draw[gp path] (6.724,1.498)--(8.330,1.498);
\gpcolor{color=gp lt color border}
\gpsetlinetype{gp lt border}
\gpsetdashtype{gp dt solid}
\gpsetlinewidth{2.00}
\draw[gp path] (6.724,1.498)--(6.814,1.498);
\draw[gp path] (8.330,1.498)--(8.240,1.498);
\node[gp node right] at (6.650,1.498) {$200$};
\gpcolor{color=gp lt color axes}
\gpsetlinetype{gp lt axes}
\gpsetdashtype{gp dt axes}
\gpsetlinewidth{0.40}
\draw[gp path] (6.724,1.961)--(8.330,1.961);
\gpcolor{color=gp lt color border}
\gpsetlinetype{gp lt border}
\gpsetdashtype{gp dt solid}
\gpsetlinewidth{2.00}
\draw[gp path] (6.724,1.961)--(6.814,1.961);
\draw[gp path] (8.330,1.961)--(8.240,1.961);
\node[gp node right] at (6.650,1.961) {$300$};
\gpcolor{color=gp lt color axes}
\gpsetlinetype{gp lt axes}
\gpsetdashtype{gp dt axes}
\gpsetlinewidth{0.40}
\draw[gp path] (7.126,0.572)--(7.126,2.332);
\gpcolor{color=gp lt color border}
\node[gp node center] at (7.126,0.326) {Busy};
\gpcolor{color=gp lt color axes}
\draw[gp path] (7.929,0.572)--(7.929,2.332);
\gpcolor{color=gp lt color border}
\node[gp node center] at (7.929,0.326) {Pause};
\gpsetlinetype{gp lt border}
\gpsetdashtype{gp dt solid}
\gpsetlinewidth{2.00}
\draw[gp path] (6.724,2.332)--(6.724,0.572)--(8.330,0.572)--(8.330,2.332)--cycle;
\node[gp node left] at (2.326,2.597) {SMT};
\node[gp node left] at (4.524,2.597) {Die};
\node[gp node left] at (6.470,2.597) {Socket};
\gpfill{rgb color={0.392,0.561,1.000}} (6.869,0.572)--(6.982,0.572)--(6.982,1.545)--(6.869,1.545)--cycle;
\gpsetlinewidth{1.00}
\draw[gp path] (6.869,0.572)--(6.869,1.544)--(6.981,1.544)--(6.981,0.572)--cycle;
\gpfill{rgb color={0.392,0.561,1.000}} (7.672,0.572)--(7.785,0.572)--(7.785,1.362)--(7.672,1.362)--cycle;
\draw[gp path] (7.672,0.572)--(7.672,1.361)--(7.784,1.361)--(7.784,0.572)--cycle;
\gpfill{rgb color={0.471,0.369,0.941}} (7.069,0.572)--(7.183,0.572)--(7.183,1.806)--(7.069,1.806)--cycle;
\draw[gp path] (7.069,0.572)--(7.069,1.805)--(7.182,1.805)--(7.182,0.572)--cycle;
\gpfill{rgb color={0.471,0.369,0.941}} (7.872,0.572)--(7.986,0.572)--(7.986,1.605)--(7.872,1.605)--cycle;
\draw[gp path] (7.872,0.572)--(7.872,1.604)--(7.985,1.604)--(7.985,0.572)--cycle;
\gpfill{rgb color={0.863,0.149,0.498}} (7.270,0.572)--(7.383,0.572)--(7.383,1.788)--(7.270,1.788)--cycle;
\draw[gp path] (7.270,0.572)--(7.270,1.787)--(7.382,1.787)--(7.382,0.572)--cycle;
\gpfill{rgb color={0.863,0.149,0.498}} (8.073,0.572)--(8.186,0.572)--(8.186,1.472)--(8.073,1.472)--cycle;
\draw[gp path] (8.073,0.572)--(8.073,1.471)--(8.185,1.471)--(8.185,0.572)--cycle;
\gpsetlinewidth{2.00}
\draw[gp path] (6.724,2.332)--(6.724,0.572)--(8.330,0.572)--(8.330,2.332)--cycle;
\node[gp node center,rotate=-270.0] at (5.902,1.452) {W};
\gpfill{rgb color={0.392,0.561,1.000}} (1.818,2.547)--(2.240,2.547)--(2.240,2.633)--(1.818,2.633)--cycle;
\gpcolor{rgb color={0.392,0.561,1.000}}
\gpsetlinewidth{1.00}
\draw[gp path] (1.818,2.547)--(1.818,2.633)--(2.240,2.633)--(2.240,2.547)--cycle;
\gpfill{rgb color={0.471,0.369,0.941}} (4.017,2.547)--(4.439,2.547)--(4.439,2.633)--(4.017,2.633)--cycle;
\gpcolor{rgb color={0.471,0.369,0.941}}
\draw[gp path] (4.017,2.547)--(4.017,2.633)--(4.439,2.633)--(4.439,2.547)--cycle;
\gpfill{rgb color={0.863,0.149,0.498}} (5.962,2.547)--(6.384,2.547)--(6.384,2.633)--(5.962,2.633)--cycle;
\gpcolor{rgb color={0.863,0.149,0.498}}
\draw[gp path] (5.962,2.547)--(5.962,2.633)--(6.384,2.633)--(6.384,2.547)--cycle;
\gpdefrectangularnode{gp plot 3}{\pgfpoint{6.724cm}{0.572cm}}{\pgfpoint{8.330cm}{2.332cm}}
\end{tikzpicture}
\caption{Per-thread scalar matrix multiplication throughput, 
  matrix multiplication core frequency, and package power 
  when co-located on the same CPU with 16 polling threads. 
  Each group compares waiting strategy (\texttt{Busy}, \texttt{Pause}) 
  with different placement strategies (Die, Socket, SMT).}%
\Description{Same as caption.}%
\label{fig:motiv:topology}%
\end{figure}

We investigate the effect of CPU topology on the
fixed budget (\autoref{fig:motiv:topology}).
Each run uses 16 threads for scalar matrix multiplication and 16 threads
for polling. We place polling threads in three topologies: applications
share dies while using different cores (Die), share the same socket but
use disjoint dies (Socket), or share a physical core with sibling
hyperthreads (SMT).
For each placement, we measure how
\texttt{Busy} and \texttt{Pause} affect
matrix multiplication throughput, frequency, and package power.

\textbf{\textit{Waiting strategy reduces interference on SMT cores.}}
When two compute-intensive applications share a physical
core, performance degrades when they compete for the same execution units.
The correct waiting strategy mitigates the slowdown from polling by freeing
the CPU power budget during idle periods.
With SMT placement, \texttt{Pause} reduces package power
by 18.7\% compared to \texttt{Busy} and improves matrix multiplication
throughput by 19.2\%.

\textbf{\textit{Thread placement can dissipate hotspots.}}
Individual chiplets can become hotspots, and DVFS can downclock
their cores. The right waiting strategy can dissipate
these hotspots so applications can run at higher clock frequencies.
For example, when \texttt{Pause} polling is co-located on 
the same die as matrix multiplication, the reduced power consumption 
on that chiplet allows the cores running matrix multiplication to 
boost from 4.28\,GHz with \texttt{Busy} polling to 4.47\,GHz. This yields
a modest 2.8\% performance improvement, but the effect is likely to be more 
pronounced in thermally constrained systems, where localised thermal
hotspots more readily limit boost frequency due to less aggressive 
cooling solutions or sustained high ambient temperatures~\cite{lu:taws}.

\par\medskip
\noindent\fbox{%
\parbox{\dimexpr\linewidth-2\fboxsep-2\fboxrule\relax}{%
\normalsize
\textbf{L2: Placement decides who shares the budget.} 
The compute budget is distributed hierarchically across cores,
chiplets, and sockets. Service placement therefore determines which
applications compete for power, frequency, and thermal headroom, not
just cache locality.
}}
\par\smallskip

\subsection{Idle Pattern Dictates Waiting Strategy}

\begin{figure}%
\centering%
\begin{tikzpicture}[gnuplot]
\tikzset{every node/.append style={font={\fontsize{8.0pt}{9.6pt}\selectfont}}}
\path (0.000,0.000) rectangle (8.458,2.997);
\gpcolor{color=gp lt color axes}
\gpsetlinetype{gp lt axes}
\gpsetdashtype{gp dt axes}
\gpsetlinewidth{0.80}
\draw[gp path] (1.310,0.574)--(7.907,0.574);
\gpcolor{color=gp lt color border}
\gpsetlinetype{gp lt border}
\gpsetdashtype{gp dt solid}
\gpsetlinewidth{2.00}
\draw[gp path] (1.310,0.574)--(1.400,0.574);
\draw[gp path] (7.907,0.574)--(7.817,0.574);
\node[gp node right] at (1.163,0.574) {$0$};
\gpcolor{color=gp lt color axes}
\gpsetlinetype{gp lt axes}
\gpsetdashtype{gp dt axes}
\gpsetlinewidth{0.80}
\draw[gp path] (1.310,1.131)--(7.907,1.131);
\gpcolor{color=gp lt color border}
\gpsetlinetype{gp lt border}
\gpsetdashtype{gp dt solid}
\gpsetlinewidth{2.00}
\draw[gp path] (1.310,1.131)--(1.400,1.131);
\draw[gp path] (7.907,1.131)--(7.817,1.131);
\node[gp node right] at (1.163,1.131) {$3000$};
\gpcolor{color=gp lt color axes}
\gpsetlinetype{gp lt axes}
\gpsetdashtype{gp dt axes}
\gpsetlinewidth{0.80}
\draw[gp path] (1.310,1.688)--(7.907,1.688);
\gpcolor{color=gp lt color border}
\gpsetlinetype{gp lt border}
\gpsetdashtype{gp dt solid}
\gpsetlinewidth{2.00}
\draw[gp path] (1.310,1.688)--(1.400,1.688);
\draw[gp path] (7.907,1.688)--(7.817,1.688);
\node[gp node right] at (1.163,1.688) {$6000$};
\gpcolor{color=gp lt color axes}
\gpsetlinetype{gp lt axes}
\gpsetdashtype{gp dt axes}
\gpsetlinewidth{0.80}
\draw[gp path] (1.310,2.245)--(7.907,2.245);
\gpcolor{color=gp lt color border}
\gpsetlinetype{gp lt border}
\gpsetdashtype{gp dt solid}
\gpsetlinewidth{2.00}
\draw[gp path] (1.310,2.245)--(1.400,2.245);
\draw[gp path] (7.907,2.245)--(7.817,2.245);
\node[gp node right] at (1.163,2.245) {$9000$};
\gpcolor{color=gp lt color axes}
\gpsetlinetype{gp lt axes}
\gpsetdashtype{gp dt axes}
\gpsetlinewidth{0.80}
\draw[gp path] (1.310,2.802)--(7.907,2.802);
\gpcolor{color=gp lt color border}
\gpsetlinetype{gp lt border}
\gpsetdashtype{gp dt solid}
\gpsetlinewidth{2.00}
\draw[gp path] (1.310,2.802)--(1.400,2.802);
\draw[gp path] (7.907,2.802)--(7.817,2.802);
\node[gp node right] at (1.163,2.802) {$12000$};
\gpcolor{color=gp lt color axes}
\gpsetlinetype{gp lt axes}
\gpsetdashtype{gp dt axes}
\gpsetlinewidth{0.80}
\draw[gp path] (2.424,0.574)--(2.424,2.802);
\gpcolor{color=gp lt color border}
\gpsetlinetype{gp lt border}
\gpsetdashtype{gp dt solid}
\gpsetlinewidth{2.00}
\draw[gp path] (2.424,0.574)--(2.424,0.664);
\node[gp node center] at (2.424,0.328) {$\mu$s};
\gpcolor{color=gp lt color axes}
\gpsetlinetype{gp lt axes}
\gpsetdashtype{gp dt axes}
\gpsetlinewidth{0.80}
\draw[gp path] (4.666,0.574)--(4.666,2.802);
\gpcolor{color=gp lt color border}
\gpsetlinetype{gp lt border}
\gpsetdashtype{gp dt solid}
\gpsetlinewidth{2.00}
\draw[gp path] (4.666,0.574)--(4.666,0.664);
\node[gp node center] at (4.666,0.328) {ms};
\gpcolor{color=gp lt color axes}
\gpsetlinetype{gp lt axes}
\gpsetdashtype{gp dt axes}
\gpsetlinewidth{0.80}
\draw[gp path] (6.908,0.574)--(6.908,2.802);
\gpcolor{color=gp lt color border}
\gpsetlinetype{gp lt border}
\gpsetdashtype{gp dt solid}
\gpsetlinewidth{2.00}
\draw[gp path] (6.908,0.574)--(6.908,0.664);
\node[gp node center] at (6.908,0.328) {s};
\draw[gp path] (1.310,2.802)--(1.310,0.574)--(7.907,0.574)--(7.907,2.802)--cycle;
\def\gpfillpath{(1.627,0.574)--(1.763,0.574)--(1.763,0.813)--(1.627,0.813)--cycle}
\gpfill{color=gpbgfillcolor} \gpfillpath;
\gpfill{rgb color={0.392,0.561,1.000},gp pattern 3,pattern color=.} \gpfillpath;
\gpcolor{rgb color={0.392,0.561,1.000}}
\gpsetlinewidth{1.00}
\draw[gp path] (1.627,0.574)--(1.627,0.812)--(1.762,0.812)--(1.762,0.574)--cycle;
\def\gpfillpath{(2.203,0.574)--(2.339,0.574)--(2.339,1.668)--(2.203,1.668)--cycle}
\gpfill{color=gpbgfillcolor} \gpfillpath;
\gpfill{rgb color={0.471,0.369,0.941},gp pattern 3,pattern color=.} \gpfillpath;
\gpcolor{rgb color={0.471,0.369,0.941}}
\draw[gp path] (2.203,0.574)--(2.203,1.667)--(2.338,1.667)--(2.338,0.574)--cycle;
\def\gpfillpath{(2.780,0.574)--(2.915,0.574)--(2.915,2.492)--(2.780,2.492)--cycle}
\gpfill{color=gpbgfillcolor} \gpfillpath;
\gpfill{rgb color={0.863,0.149,0.498},gp pattern 3,pattern color=.} \gpfillpath;
\gpcolor{rgb color={0.863,0.149,0.498}}
\draw[gp path] (2.780,0.574)--(2.780,2.491)--(2.914,2.491)--(2.914,0.574)--cycle;
\def\gpfillpath{(3.869,0.574)--(4.004,0.574)--(4.004,0.818)--(3.869,0.818)--cycle}
\gpfill{color=gpbgfillcolor} \gpfillpath;
\gpfill{rgb color={0.392,0.561,1.000},gp pattern 3,pattern color=.} \gpfillpath;
\gpcolor{rgb color={0.392,0.561,1.000}}
\draw[gp path] (3.869,0.574)--(3.869,0.817)--(4.003,0.817)--(4.003,0.574)--cycle;
\def\gpfillpath{(4.445,0.574)--(4.581,0.574)--(4.581,1.718)--(4.445,1.718)--cycle}
\gpfill{color=gpbgfillcolor} \gpfillpath;
\gpfill{rgb color={0.471,0.369,0.941},gp pattern 3,pattern color=.} \gpfillpath;
\gpcolor{rgb color={0.471,0.369,0.941}}
\draw[gp path] (4.445,0.574)--(4.445,1.717)--(4.580,1.717)--(4.580,0.574)--cycle;
\def\gpfillpath{(5.022,0.574)--(5.157,0.574)--(5.157,2.504)--(5.022,2.504)--cycle}
\gpfill{color=gpbgfillcolor} \gpfillpath;
\gpfill{rgb color={0.863,0.149,0.498},gp pattern 3,pattern color=.} \gpfillpath;
\gpcolor{rgb color={0.863,0.149,0.498}}
\draw[gp path] (5.022,0.574)--(5.022,2.503)--(5.156,2.503)--(5.156,0.574)--cycle;
\def\gpfillpath{(6.110,0.574)--(6.246,0.574)--(6.246,0.821)--(6.110,0.821)--cycle}
\gpfill{color=gpbgfillcolor} \gpfillpath;
\gpfill{rgb color={0.392,0.561,1.000},gp pattern 3,pattern color=.} \gpfillpath;
\gpcolor{rgb color={0.392,0.561,1.000}}
\draw[gp path] (6.110,0.574)--(6.110,0.820)--(6.245,0.820)--(6.245,0.574)--cycle;
\def\gpfillpath{(6.687,0.574)--(6.822,0.574)--(6.822,1.651)--(6.687,1.651)--cycle}
\gpfill{color=gpbgfillcolor} \gpfillpath;
\gpfill{rgb color={0.471,0.369,0.941},gp pattern 3,pattern color=.} \gpfillpath;
\gpcolor{rgb color={0.471,0.369,0.941}}
\draw[gp path] (6.687,0.574)--(6.687,1.650)--(6.821,1.650)--(6.821,0.574)--cycle;
\def\gpfillpath{(7.263,0.574)--(7.399,0.574)--(7.399,2.484)--(7.263,2.484)--cycle}
\gpfill{color=gpbgfillcolor} \gpfillpath;
\gpfill{rgb color={0.863,0.149,0.498},gp pattern 3,pattern color=.} \gpfillpath;
\gpcolor{rgb color={0.863,0.149,0.498}}
\draw[gp path] (7.263,0.574)--(7.263,2.483)--(7.398,2.483)--(7.398,0.574)--cycle;
\gpfill{rgb color={0.392,0.561,1.000}} (1.781,0.574)--(1.916,0.574)--(1.916,0.845)--(1.781,0.845)--cycle;
\gpcolor{rgb color={0.392,0.561,1.000}}
\draw[gp path] (1.781,0.574)--(1.781,0.844)--(1.915,0.844)--(1.915,0.574)--cycle;
\gpfill{rgb color={0.471,0.369,0.941}} (2.357,0.574)--(2.493,0.574)--(2.493,1.787)--(2.357,1.787)--cycle;
\gpcolor{rgb color={0.471,0.369,0.941}}
\draw[gp path] (2.357,0.574)--(2.357,1.786)--(2.492,1.786)--(2.492,0.574)--cycle;
\gpfill{rgb color={0.863,0.149,0.498}} (2.934,0.574)--(3.069,0.574)--(3.069,2.468)--(2.934,2.468)--cycle;
\gpcolor{rgb color={0.863,0.149,0.498}}
\draw[gp path] (2.934,0.574)--(2.934,2.467)--(3.068,2.467)--(3.068,0.574)--cycle;
\gpfill{rgb color={0.392,0.561,1.000}} (4.022,0.574)--(4.158,0.574)--(4.158,0.852)--(4.022,0.852)--cycle;
\gpcolor{rgb color={0.392,0.561,1.000}}
\draw[gp path] (4.022,0.574)--(4.022,0.851)--(4.157,0.851)--(4.157,0.574)--cycle;
\gpfill{rgb color={0.471,0.369,0.941}} (4.599,0.574)--(4.734,0.574)--(4.734,1.821)--(4.599,1.821)--cycle;
\gpcolor{rgb color={0.471,0.369,0.941}}
\draw[gp path] (4.599,0.574)--(4.599,1.820)--(4.733,1.820)--(4.733,0.574)--cycle;
\gpfill{rgb color={0.863,0.149,0.498}} (5.175,0.574)--(5.311,0.574)--(5.311,2.544)--(5.175,2.544)--cycle;
\gpcolor{rgb color={0.863,0.149,0.498}}
\draw[gp path] (5.175,0.574)--(5.175,2.543)--(5.310,2.543)--(5.310,0.574)--cycle;
\gpfill{rgb color={0.392,0.561,1.000}} (6.264,0.574)--(6.400,0.574)--(6.400,0.821)--(6.264,0.821)--cycle;
\gpcolor{rgb color={0.392,0.561,1.000}}
\draw[gp path] (6.264,0.574)--(6.264,0.820)--(6.399,0.820)--(6.399,0.574)--cycle;
\gpfill{rgb color={0.471,0.369,0.941}} (6.841,0.574)--(6.976,0.574)--(6.976,1.656)--(6.841,1.656)--cycle;
\gpcolor{rgb color={0.471,0.369,0.941}}
\draw[gp path] (6.841,0.574)--(6.841,1.655)--(6.975,1.655)--(6.975,0.574)--cycle;
\gpfill{rgb color={0.863,0.149,0.498}} (7.417,0.574)--(7.553,0.574)--(7.553,2.483)--(7.417,2.483)--cycle;
\gpcolor{rgb color={0.863,0.149,0.498}}
\draw[gp path] (7.417,0.574)--(7.417,2.482)--(7.552,2.482)--(7.552,0.574)--cycle;
\def\gpfillpath{(1.934,0.574)--(2.070,0.574)--(2.070,0.744)--(1.934,0.744)--cycle}
\gpfill{color=gpbgfillcolor} \gpfillpath;
\gpfill{rgb color={0.392,0.561,1.000},gp pattern 2,pattern color=.} \gpfillpath;
\gpcolor{rgb color={0.392,0.561,1.000}}
\draw[gp path] (1.934,0.574)--(1.934,0.743)--(2.069,0.743)--(2.069,0.574)--cycle;
\def\gpfillpath{(2.511,0.574)--(2.646,0.574)--(2.646,1.355)--(2.511,1.355)--cycle}
\gpfill{color=gpbgfillcolor} \gpfillpath;
\gpfill{rgb color={0.471,0.369,0.941},gp pattern 2,pattern color=.} \gpfillpath;
\gpcolor{rgb color={0.471,0.369,0.941}}
\draw[gp path] (2.511,0.574)--(2.511,1.354)--(2.645,1.354)--(2.645,0.574)--cycle;
\def\gpfillpath{(3.087,0.574)--(3.223,0.574)--(3.223,1.886)--(3.087,1.886)--cycle}
\gpfill{color=gpbgfillcolor} \gpfillpath;
\gpfill{rgb color={0.863,0.149,0.498},gp pattern 2,pattern color=.} \gpfillpath;
\gpcolor{rgb color={0.863,0.149,0.498}}
\draw[gp path] (3.087,0.574)--(3.087,1.885)--(3.222,1.885)--(3.222,0.574)--cycle;
\def\gpfillpath{(4.176,0.574)--(4.312,0.574)--(4.312,0.785)--(4.176,0.785)--cycle}
\gpfill{color=gpbgfillcolor} \gpfillpath;
\gpfill{rgb color={0.392,0.561,1.000},gp pattern 2,pattern color=.} \gpfillpath;
\gpcolor{rgb color={0.392,0.561,1.000}}
\draw[gp path] (4.176,0.574)--(4.176,0.784)--(4.311,0.784)--(4.311,0.574)--cycle;
\def\gpfillpath{(4.753,0.574)--(4.888,0.574)--(4.888,1.625)--(4.753,1.625)--cycle}
\gpfill{color=gpbgfillcolor} \gpfillpath;
\gpfill{rgb color={0.471,0.369,0.941},gp pattern 2,pattern color=.} \gpfillpath;
\gpcolor{rgb color={0.471,0.369,0.941}}
\draw[gp path] (4.753,0.574)--(4.753,1.624)--(4.887,1.624)--(4.887,0.574)--cycle;
\def\gpfillpath{(5.329,0.574)--(5.465,0.574)--(5.465,2.461)--(5.329,2.461)--cycle}
\gpfill{color=gpbgfillcolor} \gpfillpath;
\gpfill{rgb color={0.863,0.149,0.498},gp pattern 2,pattern color=.} \gpfillpath;
\gpcolor{rgb color={0.863,0.149,0.498}}
\draw[gp path] (5.329,0.574)--(5.329,2.460)--(5.464,2.460)--(5.464,0.574)--cycle;
\def\gpfillpath{(6.418,0.574)--(6.553,0.574)--(6.553,0.810)--(6.418,0.810)--cycle}
\gpfill{color=gpbgfillcolor} \gpfillpath;
\gpfill{rgb color={0.392,0.561,1.000},gp pattern 2,pattern color=.} \gpfillpath;
\gpcolor{rgb color={0.392,0.561,1.000}}
\draw[gp path] (6.418,0.574)--(6.418,0.809)--(6.552,0.809)--(6.552,0.574)--cycle;
\def\gpfillpath{(6.994,0.574)--(7.130,0.574)--(7.130,1.593)--(6.994,1.593)--cycle}
\gpfill{color=gpbgfillcolor} \gpfillpath;
\gpfill{rgb color={0.471,0.369,0.941},gp pattern 2,pattern color=.} \gpfillpath;
\gpcolor{rgb color={0.471,0.369,0.941}}
\draw[gp path] (6.994,0.574)--(6.994,1.592)--(7.129,1.592)--(7.129,0.574)--cycle;
\def\gpfillpath{(7.571,0.574)--(7.706,0.574)--(7.706,2.374)--(7.571,2.374)--cycle}
\gpfill{color=gpbgfillcolor} \gpfillpath;
\gpfill{rgb color={0.863,0.149,0.498},gp pattern 2,pattern color=.} \gpfillpath;
\gpcolor{rgb color={0.863,0.149,0.498}}
\draw[gp path] (7.571,0.574)--(7.571,2.373)--(7.705,2.373)--(7.705,0.574)--cycle;
\gpcolor{color=gp lt color border}
\gpsetlinewidth{1.10}
\draw[gp path] (1.694,0.812)--(1.694,0.813);
\draw[gp path] (2.271,1.666)--(2.271,1.668);
\draw[gp path] (2.847,2.487)--(2.847,2.494);
\draw[gp path] (1.848,0.844)--(1.848,0.845);
\draw[gp path] (2.424,1.779)--(2.424,1.792);
\draw[gp path] (3.001,2.463)--(3.001,2.471);
\draw[gp path] (2.578,1.353)--(2.578,1.354);
\draw[gp path] (3.155,1.881)--(3.155,1.889);
\draw[gp path] (3.936,0.814)--(3.936,0.821);
\draw[gp path] (4.512,1.714)--(4.512,1.720);
\draw[gp path] (5.089,2.501)--(5.089,2.505);
\draw[gp path] (4.090,0.850)--(4.090,0.852);
\draw[gp path] (4.666,1.811)--(4.666,1.829);
\draw[gp path] (5.243,2.541)--(5.243,2.544);
\draw[gp path] (4.243,0.783)--(4.243,0.784);
\draw[gp path] (4.820,1.623)--(4.820,1.624);
\draw[gp path] (5.396,2.451)--(5.396,2.468);
\draw[gp path] (6.178,0.819)--(6.178,0.820);
\draw[gp path] (6.754,1.649)--(6.754,1.651);
\draw[gp path] (7.331,2.479)--(7.331,2.486);
\draw[gp path] (6.908,1.649)--(6.908,1.661);
\draw[gp path] (7.484,2.480)--(7.484,2.483);
\draw[gp path] (6.485,0.808)--(6.485,0.809);
\draw[gp path] (7.062,1.592)--(7.062,1.593);
\draw[gp path] (7.638,2.371)--(7.638,2.375);
\gpsetpointsize{4.00}
\gp3point{gp mark 0}{}{(1.694,0.812)}
\gp3point{gp mark 0}{}{(2.271,1.667)}
\gp3point{gp mark 0}{}{(2.847,2.491)}
\gp3point{gp mark 0}{}{(1.848,0.844)}
\gp3point{gp mark 0}{}{(2.424,1.786)}
\gp3point{gp mark 0}{}{(3.001,2.467)}
\gp3point{gp mark 0}{}{(2.002,0.743)}
\gp3point{gp mark 0}{}{(2.578,1.354)}
\gp3point{gp mark 0}{}{(3.155,1.885)}
\gp3point{gp mark 0}{}{(3.936,0.817)}
\gp3point{gp mark 0}{}{(4.512,1.717)}
\gp3point{gp mark 0}{}{(5.089,2.503)}
\gp3point{gp mark 0}{}{(4.090,0.851)}
\gp3point{gp mark 0}{}{(4.666,1.820)}
\gp3point{gp mark 0}{}{(5.243,2.543)}
\gp3point{gp mark 0}{}{(4.243,0.784)}
\gp3point{gp mark 0}{}{(4.820,1.624)}
\gp3point{gp mark 0}{}{(5.396,2.460)}
\gp3point{gp mark 0}{}{(6.178,0.820)}
\gp3point{gp mark 0}{}{(6.754,1.650)}
\gp3point{gp mark 0}{}{(7.331,2.483)}
\gp3point{gp mark 0}{}{(6.331,0.820)}
\gp3point{gp mark 0}{}{(6.908,1.655)}
\gp3point{gp mark 0}{}{(7.484,2.482)}
\gp3point{gp mark 0}{}{(6.485,0.809)}
\gp3point{gp mark 0}{}{(7.062,1.592)}
\gp3point{gp mark 0}{}{(7.638,2.373)}
\gpsetlinewidth{2.00}
\draw[gp path] (1.310,2.802)--(1.310,0.574)--(7.907,0.574)--(7.907,2.802)--cycle;
\node[gp node center,rotate=-270.0] at (0.121,1.688) {Mops/s};
\gpdefrectangularnode{gp plot 1}{\pgfpoint{1.310cm}{0.574cm}}{\pgfpoint{7.907cm}{2.802cm}}
\end{tikzpicture}
\caption{Throughput of a bursty polling application using different
  waiting strategies,
  co-located with matrix multiplication.
  Results are grouped by burst timescale and busy:idle ratios.
  Blocking loses polling throughput at the microsecond timescale but becomes competitive at
  longer timescales. \texttt{Pause} achieves performance comparable to \texttt{Busy}
  at small timescales.}%
\Description{Same as caption.}%
\label{fig:motiv:burst-poll}%
\end{figure}
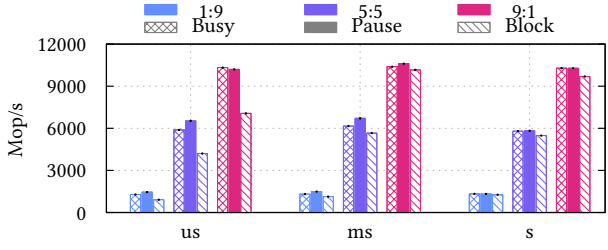

\autoref{fig:motiv:burst-app} and \autoref{fig:motiv:burst-poll}
show the trade-offs between polling strategies across burst scales
($\mu$s, ms, s) and busy:idle ratios (1:9, 5:5, 9:1). We dedicate
96 threads to matrix multiplication, while the remaining threads run a
polling application that processes headers and computes a
pseudo-checksum during busy periods, similar to network packet processing.

\begin{figure}%
\centering%
\begin{tikzpicture}[gnuplot]
\tikzset{every node/.append style={font={\fontsize{8.0pt}{9.6pt}\selectfont}}}
\path (0.000,0.000) rectangle (8.458,3.378);
\gpcolor{color=gp lt color axes}
\gpsetlinetype{gp lt axes}
\gpsetdashtype{gp dt axes}
\gpsetlinewidth{0.80}
\draw[gp path] (1.310,0.574)--(7.907,0.574);
\gpcolor{color=gp lt color border}
\gpsetlinetype{gp lt border}
\gpsetdashtype{gp dt solid}
\gpsetlinewidth{2.00}
\draw[gp path] (1.310,0.574)--(1.400,0.574);
\draw[gp path] (7.907,0.574)--(7.817,0.574);
\node[gp node right] at (1.163,0.574) {$0$};
\gpcolor{color=gp lt color axes}
\gpsetlinetype{gp lt axes}
\gpsetdashtype{gp dt axes}
\gpsetlinewidth{0.80}
\draw[gp path] (1.310,1.131)--(7.907,1.131);
\gpcolor{color=gp lt color border}
\gpsetlinetype{gp lt border}
\gpsetdashtype{gp dt solid}
\gpsetlinewidth{2.00}
\draw[gp path] (1.310,1.131)--(1.400,1.131);
\draw[gp path] (7.907,1.131)--(7.817,1.131);
\node[gp node right] at (1.163,1.131) {$0.25$};
\gpcolor{color=gp lt color axes}
\gpsetlinetype{gp lt axes}
\gpsetdashtype{gp dt axes}
\gpsetlinewidth{0.80}
\draw[gp path] (1.310,1.688)--(7.907,1.688);
\gpcolor{color=gp lt color border}
\gpsetlinetype{gp lt border}
\gpsetdashtype{gp dt solid}
\gpsetlinewidth{2.00}
\draw[gp path] (1.310,1.688)--(1.400,1.688);
\draw[gp path] (7.907,1.688)--(7.817,1.688);
\node[gp node right] at (1.163,1.688) {$0.5$};
\gpcolor{color=gp lt color axes}
\gpsetlinetype{gp lt axes}
\gpsetdashtype{gp dt axes}
\gpsetlinewidth{0.80}
\draw[gp path] (1.310,2.245)--(7.907,2.245);
\gpcolor{color=gp lt color border}
\gpsetlinetype{gp lt border}
\gpsetdashtype{gp dt solid}
\gpsetlinewidth{2.00}
\draw[gp path] (1.310,2.245)--(1.400,2.245);
\draw[gp path] (7.907,2.245)--(7.817,2.245);
\node[gp node right] at (1.163,2.245) {$0.75$};
\gpcolor{color=gp lt color axes}
\gpsetlinetype{gp lt axes}
\gpsetdashtype{gp dt axes}
\gpsetlinewidth{0.80}
\draw[gp path] (1.310,2.802)--(7.907,2.802);
\gpcolor{color=gp lt color border}
\gpsetlinetype{gp lt border}
\gpsetdashtype{gp dt solid}
\gpsetlinewidth{2.00}
\draw[gp path] (1.310,2.802)--(1.400,2.802);
\draw[gp path] (7.907,2.802)--(7.817,2.802);
\node[gp node right] at (1.163,2.802) {$1$};
\gpcolor{color=gp lt color axes}
\gpsetlinetype{gp lt axes}
\gpsetdashtype{gp dt axes}
\gpsetlinewidth{0.80}
\draw[gp path] (2.424,0.574)--(2.424,2.802);
\gpcolor{color=gp lt color border}
\gpsetlinetype{gp lt border}
\gpsetdashtype{gp dt solid}
\gpsetlinewidth{2.00}
\draw[gp path] (2.424,0.574)--(2.424,0.664);
\node[gp node center] at (2.424,0.328) {$\mu$s};
\gpcolor{color=gp lt color axes}
\gpsetlinetype{gp lt axes}
\gpsetdashtype{gp dt axes}
\gpsetlinewidth{0.80}
\draw[gp path] (4.666,0.574)--(4.666,2.802);
\gpcolor{color=gp lt color border}
\gpsetlinetype{gp lt border}
\gpsetdashtype{gp dt solid}
\gpsetlinewidth{2.00}
\draw[gp path] (4.666,0.574)--(4.666,0.664);
\node[gp node center] at (4.666,0.328) {ms};
\gpcolor{color=gp lt color axes}
\gpsetlinetype{gp lt axes}
\gpsetdashtype{gp dt axes}
\gpsetlinewidth{0.80}
\draw[gp path] (6.908,0.574)--(6.908,2.622)--(6.908,2.802);
\gpcolor{color=gp lt color border}
\gpsetlinetype{gp lt border}
\gpsetdashtype{gp dt solid}
\gpsetlinewidth{2.00}
\draw[gp path] (6.908,0.574)--(6.908,0.664);
\node[gp node center] at (6.908,0.328) {s};
\draw[gp path] (1.310,2.802)--(1.310,0.574)--(7.907,0.574)--(7.907,2.802)--cycle;
\node[gp node left] at (2.410,3.239) {1:9};
\node[gp node left] at (4.524,3.239) {5:5};
\node[gp node left] at (6.554,3.239) {9:1};
\node[gp node left] at (2.326,3.053) {Busy};
\node[gp node left] at (4.440,3.053) {Pause};
\node[gp node left] at (6.470,3.053) {Block};
\def\gpfillpath{(1.627,0.574)--(1.763,0.574)--(1.763,1.740)--(1.627,1.740)--cycle}
\gpfill{color=gpbgfillcolor} \gpfillpath;
\gpfill{rgb color={0.392,0.561,1.000},gp pattern 3,pattern color=.} \gpfillpath;
\gpcolor{rgb color={0.392,0.561,1.000}}
\gpsetlinewidth{1.00}
\draw[gp path] (1.627,0.574)--(1.627,1.739)--(1.762,1.739)--(1.762,0.574)--cycle;
\def\gpfillpath{(2.203,0.574)--(2.339,0.574)--(2.339,1.715)--(2.203,1.715)--cycle}
\gpfill{color=gpbgfillcolor} \gpfillpath;
\gpfill{rgb color={0.471,0.369,0.941},gp pattern 3,pattern color=.} \gpfillpath;
\gpcolor{rgb color={0.471,0.369,0.941}}
\draw[gp path] (2.203,0.574)--(2.203,1.714)--(2.338,1.714)--(2.338,0.574)--cycle;
\def\gpfillpath{(2.780,0.574)--(2.915,0.574)--(2.915,1.690)--(2.780,1.690)--cycle}
\gpfill{color=gpbgfillcolor} \gpfillpath;
\gpfill{rgb color={0.863,0.149,0.498},gp pattern 3,pattern color=.} \gpfillpath;
\gpcolor{rgb color={0.863,0.149,0.498}}
\draw[gp path] (2.780,0.574)--(2.780,1.689)--(2.914,1.689)--(2.914,0.574)--cycle;
\def\gpfillpath{(3.869,0.574)--(4.004,0.574)--(4.004,1.742)--(3.869,1.742)--cycle}
\gpfill{color=gpbgfillcolor} \gpfillpath;
\gpfill{rgb color={0.392,0.561,1.000},gp pattern 3,pattern color=.} \gpfillpath;
\gpcolor{rgb color={0.392,0.561,1.000}}
\draw[gp path] (3.869,0.574)--(3.869,1.741)--(4.003,1.741)--(4.003,0.574)--cycle;
\def\gpfillpath{(4.445,0.574)--(4.581,0.574)--(4.581,1.713)--(4.445,1.713)--cycle}
\gpfill{color=gpbgfillcolor} \gpfillpath;
\gpfill{rgb color={0.471,0.369,0.941},gp pattern 3,pattern color=.} \gpfillpath;
\gpcolor{rgb color={0.471,0.369,0.941}}
\draw[gp path] (4.445,0.574)--(4.445,1.712)--(4.580,1.712)--(4.580,0.574)--cycle;
\def\gpfillpath{(5.022,0.574)--(5.157,0.574)--(5.157,1.689)--(5.022,1.689)--cycle}
\gpfill{color=gpbgfillcolor} \gpfillpath;
\gpfill{rgb color={0.863,0.149,0.498},gp pattern 3,pattern color=.} \gpfillpath;
\gpcolor{rgb color={0.863,0.149,0.498}}
\draw[gp path] (5.022,0.574)--(5.022,1.688)--(5.156,1.688)--(5.156,0.574)--cycle;
\def\gpfillpath{(6.110,0.574)--(6.246,0.574)--(6.246,1.750)--(6.110,1.750)--cycle}
\gpfill{color=gpbgfillcolor} \gpfillpath;
\gpfill{rgb color={0.392,0.561,1.000},gp pattern 3,pattern color=.} \gpfillpath;
\gpcolor{rgb color={0.392,0.561,1.000}}
\draw[gp path] (6.110,0.574)--(6.110,1.749)--(6.245,1.749)--(6.245,0.574)--cycle;
\def\gpfillpath{(6.687,0.574)--(6.822,0.574)--(6.822,1.722)--(6.687,1.722)--cycle}
\gpfill{color=gpbgfillcolor} \gpfillpath;
\gpfill{rgb color={0.471,0.369,0.941},gp pattern 3,pattern color=.} \gpfillpath;
\gpcolor{rgb color={0.471,0.369,0.941}}
\draw[gp path] (6.687,0.574)--(6.687,1.721)--(6.821,1.721)--(6.821,0.574)--cycle;
\def\gpfillpath{(7.263,0.574)--(7.399,0.574)--(7.399,1.692)--(7.263,1.692)--cycle}
\gpfill{color=gpbgfillcolor} \gpfillpath;
\gpfill{rgb color={0.863,0.149,0.498},gp pattern 3,pattern color=.} \gpfillpath;
\gpcolor{rgb color={0.863,0.149,0.498}}
\draw[gp path] (7.263,0.574)--(7.263,1.691)--(7.398,1.691)--(7.398,0.574)--cycle;
\gpfill{rgb color={0.392,0.561,1.000}} (1.781,0.574)--(1.916,0.574)--(1.916,1.820)--(1.781,1.820)--cycle;
\gpcolor{rgb color={0.392,0.561,1.000}}
\draw[gp path] (1.781,0.574)--(1.781,1.819)--(1.915,1.819)--(1.915,0.574)--cycle;
\gpfill{rgb color={0.471,0.369,0.941}} (2.357,0.574)--(2.493,0.574)--(2.493,1.761)--(2.357,1.761)--cycle;
\gpcolor{rgb color={0.471,0.369,0.941}}
\draw[gp path] (2.357,0.574)--(2.357,1.760)--(2.492,1.760)--(2.492,0.574)--cycle;
\gpfill{rgb color={0.863,0.149,0.498}} (2.934,0.574)--(3.069,0.574)--(3.069,1.703)--(2.934,1.703)--cycle;
\gpcolor{rgb color={0.863,0.149,0.498}}
\draw[gp path] (2.934,0.574)--(2.934,1.702)--(3.068,1.702)--(3.068,0.574)--cycle;
\gpfill{rgb color={0.392,0.561,1.000}} (4.022,0.574)--(4.158,0.574)--(4.158,1.808)--(4.022,1.808)--cycle;
\gpcolor{rgb color={0.392,0.561,1.000}}
\draw[gp path] (4.022,0.574)--(4.022,1.807)--(4.157,1.807)--(4.157,0.574)--cycle;
\gpfill{rgb color={0.471,0.369,0.941}} (4.599,0.574)--(4.734,0.574)--(4.734,1.758)--(4.599,1.758)--cycle;
\gpcolor{rgb color={0.471,0.369,0.941}}
\draw[gp path] (4.599,0.574)--(4.599,1.757)--(4.733,1.757)--(4.733,0.574)--cycle;
\gpfill{rgb color={0.863,0.149,0.498}} (5.175,0.574)--(5.311,0.574)--(5.311,1.696)--(5.175,1.696)--cycle;
\gpcolor{rgb color={0.863,0.149,0.498}}
\draw[gp path] (5.175,0.574)--(5.175,1.695)--(5.310,1.695)--(5.310,0.574)--cycle;
\gpfill{rgb color={0.392,0.561,1.000}} (6.264,0.574)--(6.400,0.574)--(6.400,1.823)--(6.264,1.823)--cycle;
\gpcolor{rgb color={0.392,0.561,1.000}}
\draw[gp path] (6.264,0.574)--(6.264,1.822)--(6.399,1.822)--(6.399,0.574)--cycle;
\gpfill{rgb color={0.471,0.369,0.941}} (6.841,0.574)--(6.976,0.574)--(6.976,1.765)--(6.841,1.765)--cycle;
\gpcolor{rgb color={0.471,0.369,0.941}}
\draw[gp path] (6.841,0.574)--(6.841,1.764)--(6.975,1.764)--(6.975,0.574)--cycle;
\gpfill{rgb color={0.863,0.149,0.498}} (7.417,0.574)--(7.553,0.574)--(7.553,1.699)--(7.417,1.699)--cycle;
\gpcolor{rgb color={0.863,0.149,0.498}}
\draw[gp path] (7.417,0.574)--(7.417,1.698)--(7.552,1.698)--(7.552,0.574)--cycle;
\def\gpfillpath{(1.934,0.574)--(2.070,0.574)--(2.070,2.179)--(1.934,2.179)--cycle}
\gpfill{color=gpbgfillcolor} \gpfillpath;
\gpfill{rgb color={0.392,0.561,1.000},gp pattern 2,pattern color=.} \gpfillpath;
\gpcolor{rgb color={0.392,0.561,1.000}}
\draw[gp path] (1.934,0.574)--(1.934,2.178)--(2.069,2.178)--(2.069,0.574)--cycle;
\def\gpfillpath{(2.511,0.574)--(2.646,0.574)--(2.646,1.902)--(2.511,1.902)--cycle}
\gpfill{color=gpbgfillcolor} \gpfillpath;
\gpfill{rgb color={0.471,0.369,0.941},gp pattern 2,pattern color=.} \gpfillpath;
\gpcolor{rgb color={0.471,0.369,0.941}}
\draw[gp path] (2.511,0.574)--(2.511,1.901)--(2.645,1.901)--(2.645,0.574)--cycle;
\def\gpfillpath{(3.087,0.574)--(3.223,0.574)--(3.223,1.749)--(3.087,1.749)--cycle}
\gpfill{color=gpbgfillcolor} \gpfillpath;
\gpfill{rgb color={0.863,0.149,0.498},gp pattern 2,pattern color=.} \gpfillpath;
\gpcolor{rgb color={0.863,0.149,0.498}}
\draw[gp path] (3.087,0.574)--(3.087,1.748)--(3.222,1.748)--(3.222,0.574)--cycle;
\def\gpfillpath{(4.176,0.574)--(4.312,0.574)--(4.312,2.691)--(4.176,2.691)--cycle}
\gpfill{color=gpbgfillcolor} \gpfillpath;
\gpfill{rgb color={0.392,0.561,1.000},gp pattern 2,pattern color=.} \gpfillpath;
\gpcolor{rgb color={0.392,0.561,1.000}}
\draw[gp path] (4.176,0.574)--(4.176,2.690)--(4.311,2.690)--(4.311,0.574)--cycle;
\def\gpfillpath{(4.753,0.574)--(4.888,0.574)--(4.888,2.252)--(4.753,2.252)--cycle}
\gpfill{color=gpbgfillcolor} \gpfillpath;
\gpfill{rgb color={0.471,0.369,0.941},gp pattern 2,pattern color=.} \gpfillpath;
\gpcolor{rgb color={0.471,0.369,0.941}}
\draw[gp path] (4.753,0.574)--(4.753,2.251)--(4.887,2.251)--(4.887,0.574)--cycle;
\def\gpfillpath{(5.329,0.574)--(5.465,0.574)--(5.465,1.814)--(5.329,1.814)--cycle}
\gpfill{color=gpbgfillcolor} \gpfillpath;
\gpfill{rgb color={0.863,0.149,0.498},gp pattern 2,pattern color=.} \gpfillpath;
\gpcolor{rgb color={0.863,0.149,0.498}}
\draw[gp path] (5.329,0.574)--(5.329,1.813)--(5.464,1.813)--(5.464,0.574)--cycle;
\def\gpfillpath{(6.418,0.574)--(6.553,0.574)--(6.553,2.678)--(6.418,2.678)--cycle}
\gpfill{color=gpbgfillcolor} \gpfillpath;
\gpfill{rgb color={0.392,0.561,1.000},gp pattern 2,pattern color=.} \gpfillpath;
\gpcolor{rgb color={0.392,0.561,1.000}}
\draw[gp path] (6.418,0.574)--(6.418,2.677)--(6.552,2.677)--(6.552,0.574)--cycle;
\def\gpfillpath{(6.994,0.574)--(7.130,0.574)--(7.130,2.268)--(6.994,2.268)--cycle}
\gpfill{color=gpbgfillcolor} \gpfillpath;
\gpfill{rgb color={0.471,0.369,0.941},gp pattern 2,pattern color=.} \gpfillpath;
\gpcolor{rgb color={0.471,0.369,0.941}}
\draw[gp path] (6.994,0.574)--(6.994,2.267)--(7.129,2.267)--(7.129,0.574)--cycle;
\def\gpfillpath{(7.571,0.574)--(7.706,0.574)--(7.706,1.864)--(7.571,1.864)--cycle}
\gpfill{color=gpbgfillcolor} \gpfillpath;
\gpfill{rgb color={0.863,0.149,0.498},gp pattern 2,pattern color=.} \gpfillpath;
\gpcolor{rgb color={0.863,0.149,0.498}}
\draw[gp path] (7.571,0.574)--(7.571,1.863)--(7.705,1.863)--(7.705,0.574)--cycle;
\gpcolor{color=gp lt color border}
\gpsetlinewidth{1.10}
\draw[gp path] (1.694,1.739)--(1.694,1.740);
\draw[gp path] (2.271,1.712)--(2.271,1.716);
\draw[gp path] (2.847,1.688)--(2.847,1.690);
\draw[gp path] (1.848,1.818)--(1.848,1.819);
\draw[gp path] (2.424,1.758)--(2.424,1.762);
\draw[gp path] (3.001,1.700)--(3.001,1.704);
\draw[gp path] (2.002,2.175)--(2.002,2.181);
\draw[gp path] (2.578,1.897)--(2.578,1.905);
\draw[gp path] (3.155,1.747)--(3.155,1.749);
\draw[gp path] (3.936,1.739)--(3.936,1.742);
\draw[gp path] (4.512,1.710)--(4.512,1.713);
\draw[gp path] (5.089,1.687)--(5.089,1.689);
\draw[gp path] (4.090,1.804)--(4.090,1.811);
\draw[gp path] (4.666,1.756)--(4.666,1.758);
\draw[gp path] (5.243,1.695)--(5.243,1.696);
\draw[gp path] (4.243,2.688)--(4.243,2.691);
\draw[gp path] (4.820,2.249)--(4.820,2.252);
\draw[gp path] (5.396,1.812)--(5.396,1.814);
\draw[gp path] (6.178,1.748)--(6.178,1.750);
\draw[gp path] (6.754,1.718)--(6.754,1.724);
\draw[gp path] (7.331,1.690)--(7.331,1.693);
\draw[gp path] (6.908,1.758)--(6.908,1.770);
\draw[gp path] (7.484,1.697)--(7.484,1.699);
\draw[gp path] (6.485,2.673)--(6.485,2.680);
\draw[gp path] (7.062,2.266)--(7.062,2.268);
\draw[gp path] (7.638,1.862)--(7.638,1.864);
\gpsetpointsize{4.00}
\gp3point{gp mark 0}{}{(1.694,1.739)}
\gp3point{gp mark 0}{}{(2.271,1.714)}
\gp3point{gp mark 0}{}{(2.847,1.689)}
\gp3point{gp mark 0}{}{(1.848,1.819)}
\gp3point{gp mark 0}{}{(2.424,1.760)}
\gp3point{gp mark 0}{}{(3.001,1.702)}
\gp3point{gp mark 0}{}{(2.002,2.178)}
\gp3point{gp mark 0}{}{(2.578,1.901)}
\gp3point{gp mark 0}{}{(3.155,1.748)}
\gp3point{gp mark 0}{}{(3.936,1.741)}
\gp3point{gp mark 0}{}{(4.512,1.712)}
\gp3point{gp mark 0}{}{(5.089,1.688)}
\gp3point{gp mark 0}{}{(4.090,1.807)}
\gp3point{gp mark 0}{}{(4.666,1.757)}
\gp3point{gp mark 0}{}{(5.243,1.695)}
\gp3point{gp mark 0}{}{(4.243,2.690)}
\gp3point{gp mark 0}{}{(4.820,2.251)}
\gp3point{gp mark 0}{}{(5.396,1.813)}
\gp3point{gp mark 0}{}{(6.178,1.749)}
\gp3point{gp mark 0}{}{(6.754,1.721)}
\gp3point{gp mark 0}{}{(7.331,1.691)}
\gp3point{gp mark 0}{}{(6.331,1.822)}
\gp3point{gp mark 0}{}{(6.908,1.764)}
\gp3point{gp mark 0}{}{(7.484,1.698)}
\gp3point{gp mark 0}{}{(6.485,2.677)}
\gp3point{gp mark 0}{}{(7.062,2.267)}
\gp3point{gp mark 0}{}{(7.638,1.863)}
\gpsetlinewidth{2.00}
\draw[gp path] (1.310,2.802)--(1.310,0.574)--(7.907,0.574)--(7.907,2.802)--cycle;
\node[gp node center,rotate=-270.0] at (0.268,1.688) {Norm Ops/s};
\gpfill{rgb color={0.392,0.561,1.000}} (1.818,3.191)--(2.240,3.191)--(2.240,3.326)--(1.818,3.326)--cycle;
\gpcolor{rgb color={0.392,0.561,1.000}}
\gpsetlinewidth{1.00}
\draw[gp path] (1.818,3.191)--(1.818,3.326)--(2.240,3.326)--(2.240,3.191)--cycle;
\gpfill{rgb color={0.471,0.369,0.941}} (3.933,3.191)--(4.355,3.191)--(4.355,3.326)--(3.933,3.326)--cycle;
\gpcolor{rgb color={0.471,0.369,0.941}}
\draw[gp path] (3.933,3.191)--(3.933,3.326)--(4.355,3.326)--(4.355,3.191)--cycle;
\gpfill{rgb color={0.863,0.149,0.498}} (5.962,3.191)--(6.384,3.191)--(6.384,3.326)--(5.962,3.326)--cycle;
\gpcolor{rgb color={0.863,0.149,0.498}}
\draw[gp path] (5.962,3.191)--(5.962,3.326)--(6.384,3.326)--(6.384,3.191)--cycle;
\def\gpfillpath{(1.818,2.955)--(2.240,2.955)--(2.240,3.090)--(1.818,3.090)--cycle}
\gpfill{color=gpbgfillcolor} \gpfillpath;
\gpfill{rgb color={0.502,0.502,0.502},gp pattern 3,pattern color=.} \gpfillpath;
\gpcolor{rgb color={0.502,0.502,0.502}}
\draw[gp path] (1.818,2.955)--(1.818,3.090)--(2.240,3.090)--(2.240,2.955)--cycle;
\gpfill{rgb color={0.502,0.502,0.502}} (3.933,2.955)--(4.355,2.955)--(4.355,3.090)--(3.933,3.090)--cycle;
\draw[gp path] (3.933,2.955)--(3.933,3.090)--(4.355,3.090)--(4.355,2.955)--cycle;
\def\gpfillpath{(5.962,2.955)--(6.384,2.955)--(6.384,3.090)--(5.962,3.090)--cycle}
\gpfill{color=gpbgfillcolor} \gpfillpath;
\gpfill{rgb color={0.502,0.502,0.502},gp pattern 2,pattern color=.} \gpfillpath;
\draw[gp path] (5.962,2.955)--(5.962,3.090)--(6.384,3.090)--(6.384,2.955)--cycle;
\gpdefrectangularnode{gp plot 1}{\pgfpoint{1.310cm}{0.574cm}}{\pgfpoint{7.907cm}{2.802cm}}
\end{tikzpicture}
\caption{Normalised scalar matrix multiplication throughput with
  co-located bursty polling workloads using different
  waiting strategies (\texttt{Busy}, \texttt{Pause}, and \texttt{Block}).
  Results are grouped by burst timescale (\(\mu\)s, ms, and s) and
  busy:idle ratios (1:9, 5:5, and 9:1).
  Blocking frees the most application capacity when
  idle periods are long enough to amortise scheduler overhead, but \texttt{Pause}
  helps application capacity at short timescales.}%
\Description{Same as caption.}%
\label{fig:motiv:burst-app}%
\end{figure}
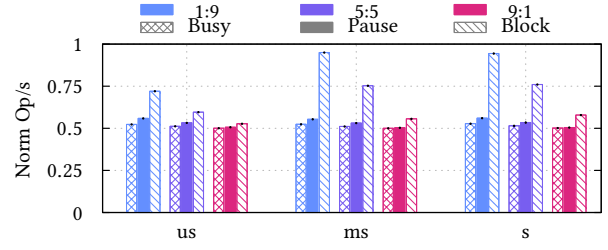

We report both applications' aggregate throughput for each
strategy. With \texttt{Block}, polling threads run with real-time
priority and share logical cores with additional matrix multiplication
threads. Sleeping polling threads yield to matrix multiplication, and
waking polling threads immediately preempt them.
\autoref{fig:motiv:burst-app} normalises throughput to matrix multiplication
running alone on all 192 threads.

\textbf{\textit{Waiting strategy impacts application performance.}}
Matrix multiplication benefits from blocking polling during idle
periods. At the second timescale,
blocking reduces throughput by only 5.7\% from the baseline when the polling
application is busy for one second and idle for nine. At the
microsecond timescale, the overheads are not amortised and blocking
incurs a 28.1\% slowdown relative to the baseline.
In comparison, we measure a 47.7\% slowdown for busy polling at the microsecond
timescale. \texttt{Pause} polling does slightly better, at 44.2\% slowdown.

\textbf{\textit{Wakeup overheads impact polling performance.}}
The overheads and wakeup latencies of the different strategies
influence polling performance. At the microsecond timescale,
blocking processes 0.9 million pseudo-packets per second for a 1:9
busy:idle ratio. Polling with \texttt{Pause}, in contrast,
processes 1.46 million pseudo-packets per second, a 62.2\% speedup.
At the second timescale, scheduler overheads are amortised and
blocking achieves performance comparable to busy and
\texttt{Pause} polling.

\par\medskip
\noindent\fbox{%
\parbox{\dimexpr\linewidth-2\fboxsep-2\fboxrule\relax}{%
\normalsize
\textbf{L3: Waiting policy should be adaptive.} 
Waiting spans a spectrum between busy polling and blocking. Different
points trade wakeup latency for returned compute budget. Network
stacks should adapt their waiting mechanism to expected idle duration
instead of treating polling and blocking as fixed alternatives.
}}
\par\smallskip

\par\medskip
\noindent\fbox{%
\parbox{\dimexpr\linewidth-2\fboxsep-2\fboxrule\relax}{%
\normalsize
\textbf{L4: Avoid complexity at small timescales.} 
Core reallocation is worthwhile only after scheduler overheads,
migration costs, and cache disruption can be amortised. At
microsecond timescales, retaining ownership and selecting a better
waiting policy to redistribute the fixed budget is often more 
effective than moving work between cores.
}}
\par\smallskip

\subsection{Downclocking Network Stack Cores Boosts RPC Throughput}
The preceding microbenchmarks show how waiting policies and thread
placement affect the budget available to productive work.
We now test whether the same model explains end-to-end RPC performance
with a real service-core network stack.
We also examine a second way to redistribute the budget: explicitly
reducing network-core frequency, rather than changing how those cores
wait.

\autoref{fig:motiv:network} illustrates this on two AWS
\texttt{c8a.metal-24xl} bare-metal instances, each with 96 cores (AMD
EPYC~9R45).
Each host runs Chamelio~\cite{stolet:chamelio} as its service-core
network stack with 48 fast-path cores, and an RPC server application
on 48 separate cores.
These bare-metal AWS instances are capped at 320\,W,
so the effects observed here reflect redistribution of power headroom.
The client sends fixed 64-byte UDP RPC 
requests at a constant configured rate.

These AWS instances limit software frequency control to two groups of 48 cores.
We therefore place the network datapath and application workers in separate 
groups to vary datapath frequency independently. 
This restriction comes from the AWS host control interface; 
AMD documents per-core boost limits through HSMP, 
which is unavailable here.

\textbf{\textit{Slowing down network cores increases throughput.}}
In the graph to the left, 
the server performs 20,000 arithmetic operations per 
request to model a compute-intensive service.
Reducing the server's fast-path
frequency from 4.5 to 2.12\,GHz increases throughput from 1.60 to
1.84 million requests per second, a 14.6\% improvement. Throughput remains
near 1.60 million requests per second between 3.71 and 4.5\,GHz,
then rises as fast-path frequency decreases further. The management unit
downclocks the whole package when the CPU hits its power cap.
Selectively reducing the frequency of the networking stack gives
power headroom to application cores and prevents them from being excessively downclocked. 
This result is consistent with the fixed-budget model: reducing the fast path's demand
for the processor's budget leaves more budget for application workers.
The throughput measurements show that maximising fast-path frequency can
reduce end-to-end performance.

\textbf{\textit{Throughput gains grow with operations per request.}}
In the graph to the right, we vary the amount of work per request. 
We measure the throughput gain from running Chamelio’s 
fast-path cores at 2.12\,GHz instead of 4.5\,GHz.
The gain from reducing fast-path frequency grows with server work per request.
At 1 Kop/req, lower frequencies improve throughput by
4.4\%. At 5 Kop/req, the improvement is 9.0\%. It reaches 13.5\% at
10 Kop/req and about 14\% at 15 and 20 Kop/req. As work per request grows,
the application benefits more from the headroom released by
the network stack.

\begin{figure}%
\centering%
\begin{tikzpicture}[gnuplot]
\tikzset{every node/.append style={font={\fontsize{8.0pt}{9.6pt}\selectfont}}}
\path (0.000,0.000) rectangle (8.458,3.378);
\gpcolor{color=gp lt color axes}
\gpsetlinetype{gp lt axes}
\gpsetdashtype{gp dt axes}
\gpsetlinewidth{0.80}
\draw[gp path] (1.057,0.675)--(4.017,0.675);
\gpcolor{color=gp lt color border}
\gpsetlinetype{gp lt border}
\gpsetdashtype{gp dt solid}
\gpsetlinewidth{2.00}
\draw[gp path] (1.057,0.675)--(1.147,0.675);
\draw[gp path] (4.017,0.675)--(3.927,0.675);
\node[gp node right] at (0.983,0.675) {$0$};
\gpcolor{color=gp lt color axes}
\gpsetlinetype{gp lt axes}
\gpsetdashtype{gp dt axes}
\gpsetlinewidth{0.80}
\draw[gp path] (1.057,1.903)--(4.017,1.903);
\gpcolor{color=gp lt color border}
\gpsetlinetype{gp lt border}
\gpsetdashtype{gp dt solid}
\gpsetlinewidth{2.00}
\draw[gp path] (1.057,1.903)--(1.147,1.903);
\draw[gp path] (4.017,1.903)--(3.927,1.903);
\node[gp node right] at (0.983,1.903) {$1$};
\gpcolor{color=gp lt color axes}
\gpsetlinetype{gp lt axes}
\gpsetdashtype{gp dt axes}
\gpsetlinewidth{0.80}
\draw[gp path] (1.057,3.131)--(4.017,3.131);
\gpcolor{color=gp lt color border}
\gpsetlinetype{gp lt border}
\gpsetdashtype{gp dt solid}
\gpsetlinewidth{2.00}
\draw[gp path] (1.057,3.131)--(1.147,3.131);
\draw[gp path] (4.017,3.131)--(3.927,3.131);
\node[gp node right] at (0.983,3.131) {$2$};
\gpcolor{color=gp lt color axes}
\gpsetlinetype{gp lt axes}
\gpsetdashtype{gp dt axes}
\gpsetlinewidth{0.80}
\draw[gp path] (1.057,0.675)--(1.057,3.131);
\gpcolor{color=gp lt color border}
\gpsetlinetype{gp lt border}
\gpsetdashtype{gp dt solid}
\gpsetlinewidth{2.00}
\draw[gp path] (1.057,0.675)--(1.057,0.765);
\node[gp node center] at (1.057,0.429) {$2$};
\gpcolor{color=gp lt color axes}
\gpsetlinetype{gp lt axes}
\gpsetdashtype{gp dt axes}
\gpsetlinewidth{0.80}
\draw[gp path] (1.626,0.675)--(1.626,3.131);
\gpcolor{color=gp lt color border}
\gpsetlinetype{gp lt border}
\gpsetdashtype{gp dt solid}
\gpsetlinewidth{2.00}
\draw[gp path] (1.626,0.675)--(1.626,0.765);
\node[gp node center] at (1.626,0.429) {$2.5$};
\gpcolor{color=gp lt color axes}
\gpsetlinetype{gp lt axes}
\gpsetdashtype{gp dt axes}
\gpsetlinewidth{0.80}
\draw[gp path] (2.195,0.675)--(2.195,3.131);
\gpcolor{color=gp lt color border}
\gpsetlinetype{gp lt border}
\gpsetdashtype{gp dt solid}
\gpsetlinewidth{2.00}
\draw[gp path] (2.195,0.675)--(2.195,0.765);
\node[gp node center] at (2.195,0.429) {$3$};
\gpcolor{color=gp lt color axes}
\gpsetlinetype{gp lt axes}
\gpsetdashtype{gp dt axes}
\gpsetlinewidth{0.80}
\draw[gp path] (2.765,0.675)--(2.765,3.131);
\gpcolor{color=gp lt color border}
\gpsetlinetype{gp lt border}
\gpsetdashtype{gp dt solid}
\gpsetlinewidth{2.00}
\draw[gp path] (2.765,0.675)--(2.765,0.765);
\node[gp node center] at (2.765,0.429) {$3.5$};
\gpcolor{color=gp lt color axes}
\gpsetlinetype{gp lt axes}
\gpsetdashtype{gp dt axes}
\gpsetlinewidth{0.80}
\draw[gp path] (3.334,0.675)--(3.334,3.131);
\gpcolor{color=gp lt color border}
\gpsetlinetype{gp lt border}
\gpsetdashtype{gp dt solid}
\gpsetlinewidth{2.00}
\draw[gp path] (3.334,0.675)--(3.334,0.765);
\node[gp node center] at (3.334,0.429) {$4$};
\gpcolor{color=gp lt color axes}
\gpsetlinetype{gp lt axes}
\gpsetdashtype{gp dt axes}
\gpsetlinewidth{0.80}
\draw[gp path] (3.903,0.675)--(3.903,3.131);
\gpcolor{color=gp lt color border}
\gpsetlinetype{gp lt border}
\gpsetdashtype{gp dt solid}
\gpsetlinewidth{2.00}
\draw[gp path] (3.903,0.675)--(3.903,0.765);
\node[gp node center] at (3.903,0.429) {$4.5$};
\draw[gp path] (1.057,3.131)--(1.057,0.675)--(4.017,0.675)--(4.017,3.131)--cycle;
\gpcolor{rgb color={0.863,0.149,0.498}}
\draw[gp path] (3.903,2.642)--(3.688,2.643)--(3.461,2.645)--(3.234,2.640)--(3.007,2.639)%
  --(2.781,2.711)--(2.553,2.783)--(2.326,2.828)--(2.100,2.862)--(1.872,2.875)--(1.646,2.888)%
  --(1.419,2.919)--(1.191,2.929);
\gpsetpointsize{4.00}
\gp3point{gp mark 7}{}{(3.903,2.642)}
\gp3point{gp mark 7}{}{(3.688,2.643)}
\gp3point{gp mark 7}{}{(3.461,2.645)}
\gp3point{gp mark 7}{}{(3.234,2.640)}
\gp3point{gp mark 7}{}{(3.007,2.639)}
\gp3point{gp mark 7}{}{(2.781,2.711)}
\gp3point{gp mark 7}{}{(2.553,2.783)}
\gp3point{gp mark 7}{}{(2.326,2.828)}
\gp3point{gp mark 7}{}{(2.100,2.862)}
\gp3point{gp mark 7}{}{(1.872,2.875)}
\gp3point{gp mark 7}{}{(1.646,2.888)}
\gp3point{gp mark 7}{}{(1.419,2.919)}
\gp3point{gp mark 7}{}{(1.191,2.929)}
\gpcolor{color=gp lt color border}
\draw[gp path] (1.057,3.131)--(1.057,0.675)--(4.017,0.675)--(4.017,3.131)--cycle;
\node[gp node center,rotate=-270.0] at (0.529,1.903) {Tput [Mreq/s]};
\node[gp node center] at (2.537,0.060) {Frequency [GHz]};
\gpdefrectangularnode{gp plot 1}{\pgfpoint{1.057cm}{0.675cm}}{\pgfpoint{4.017cm}{3.131cm}}
\gpcolor{color=gp lt color axes}
\gpsetlinetype{gp lt axes}
\gpsetdashtype{gp dt axes}
\gpsetlinewidth{0.80}
\draw[gp path] (5.201,0.675)--(8.118,0.675);
\gpcolor{color=gp lt color border}
\gpsetlinetype{gp lt border}
\gpsetdashtype{gp dt solid}
\gpsetlinewidth{2.00}
\draw[gp path] (5.201,0.675)--(5.291,0.675);
\draw[gp path] (8.118,0.675)--(8.028,0.675);
\node[gp node right] at (5.127,0.675) {$0$};
\gpcolor{color=gp lt color axes}
\gpsetlinetype{gp lt axes}
\gpsetdashtype{gp dt axes}
\gpsetlinewidth{0.80}
\draw[gp path] (5.201,1.494)--(8.118,1.494);
\gpcolor{color=gp lt color border}
\gpsetlinetype{gp lt border}
\gpsetdashtype{gp dt solid}
\gpsetlinewidth{2.00}
\draw[gp path] (5.201,1.494)--(5.291,1.494);
\draw[gp path] (8.118,1.494)--(8.028,1.494);
\node[gp node right] at (5.127,1.494) {$5$};
\gpcolor{color=gp lt color axes}
\gpsetlinetype{gp lt axes}
\gpsetdashtype{gp dt axes}
\gpsetlinewidth{0.80}
\draw[gp path] (5.201,2.312)--(8.118,2.312);
\gpcolor{color=gp lt color border}
\gpsetlinetype{gp lt border}
\gpsetdashtype{gp dt solid}
\gpsetlinewidth{2.00}
\draw[gp path] (5.201,2.312)--(5.291,2.312);
\draw[gp path] (8.118,2.312)--(8.028,2.312);
\node[gp node right] at (5.127,2.312) {$10$};
\gpcolor{color=gp lt color axes}
\gpsetlinetype{gp lt axes}
\gpsetdashtype{gp dt axes}
\gpsetlinewidth{0.80}
\draw[gp path] (5.201,3.131)--(8.118,3.131);
\gpcolor{color=gp lt color border}
\gpsetlinetype{gp lt border}
\gpsetdashtype{gp dt solid}
\gpsetlinewidth{2.00}
\draw[gp path] (5.201,3.131)--(5.291,3.131);
\draw[gp path] (8.118,3.131)--(8.028,3.131);
\node[gp node right] at (5.127,3.131) {$15$};
\gpcolor{color=gp lt color axes}
\gpsetlinetype{gp lt axes}
\gpsetdashtype{gp dt axes}
\gpsetlinewidth{0.80}
\draw[gp path] (5.201,0.675)--(5.201,3.131);
\gpcolor{color=gp lt color border}
\gpsetlinetype{gp lt border}
\gpsetdashtype{gp dt solid}
\gpsetlinewidth{2.00}
\draw[gp path] (5.201,0.675)--(5.201,0.765);
\node[gp node center] at (5.201,0.429) {$0$};
\gpcolor{color=gp lt color axes}
\gpsetlinetype{gp lt axes}
\gpsetdashtype{gp dt axes}
\gpsetlinewidth{0.80}
\draw[gp path] (5.896,0.675)--(5.896,3.131);
\gpcolor{color=gp lt color border}
\gpsetlinetype{gp lt border}
\gpsetdashtype{gp dt solid}
\gpsetlinewidth{2.00}
\draw[gp path] (5.896,0.675)--(5.896,0.765);
\node[gp node center] at (5.896,0.429) {$5$};
\gpcolor{color=gp lt color axes}
\gpsetlinetype{gp lt axes}
\gpsetdashtype{gp dt axes}
\gpsetlinewidth{0.80}
\draw[gp path] (6.590,0.675)--(6.590,3.131);
\gpcolor{color=gp lt color border}
\gpsetlinetype{gp lt border}
\gpsetdashtype{gp dt solid}
\gpsetlinewidth{2.00}
\draw[gp path] (6.590,0.675)--(6.590,0.765);
\node[gp node center] at (6.590,0.429) {$10$};
\gpcolor{color=gp lt color axes}
\gpsetlinetype{gp lt axes}
\gpsetdashtype{gp dt axes}
\gpsetlinewidth{0.80}
\draw[gp path] (7.285,0.675)--(7.285,3.131);
\gpcolor{color=gp lt color border}
\gpsetlinetype{gp lt border}
\gpsetdashtype{gp dt solid}
\gpsetlinewidth{2.00}
\draw[gp path] (7.285,0.675)--(7.285,0.765);
\node[gp node center] at (7.285,0.429) {$15$};
\gpcolor{color=gp lt color axes}
\gpsetlinetype{gp lt axes}
\gpsetdashtype{gp dt axes}
\gpsetlinewidth{0.80}
\draw[gp path] (7.979,0.675)--(7.979,3.131);
\gpcolor{color=gp lt color border}
\gpsetlinetype{gp lt border}
\gpsetdashtype{gp dt solid}
\gpsetlinewidth{2.00}
\draw[gp path] (7.979,0.675)--(7.979,0.765);
\node[gp node center] at (7.979,0.429) {$20$};
\draw[gp path] (5.201,3.131)--(5.201,0.675)--(8.118,0.675)--(8.118,3.131)--cycle;
\gpcolor{rgb color={0.863,0.149,0.498}}
\draw[gp path] (5.340,1.391)--(5.896,2.145)--(6.590,2.888)--(7.285,2.914)--(7.979,3.027);
\gp3point{gp mark 7}{}{(5.340,1.391)}
\gp3point{gp mark 7}{}{(5.896,2.145)}
\gp3point{gp mark 7}{}{(6.590,2.888)}
\gp3point{gp mark 7}{}{(7.285,2.914)}
\gp3point{gp mark 7}{}{(7.979,3.027)}
\gpcolor{color=gp lt color border}
\draw[gp path] (5.201,3.131)--(5.201,0.675)--(8.118,0.675)--(8.118,3.131)--cycle;
\node[gp node center,rotate=-270.0] at (4.526,1.903) {Tput Gain [\%]};
\node[gp node center] at (6.659,0.060) {Kop/req};
\gpdefrectangularnode{gp plot 2}{\pgfpoint{5.201cm}{0.675cm}}{\pgfpoint{8.118cm}{3.131cm}}
\end{tikzpicture}
\caption{RPC throughput vs. network stack service core frequency (left) and
  gains from reducing service core frequency with varying work per request (right).
  Slowing the network stack releases package budget to application cores and improves
  throughput. More application work per request increases the benefit.}%
\Description{Same as caption.}%
\label{fig:motiv:network}%
\end{figure}

These results extend the fixed-budget model to a real network stack:
reducing the service cores’ budget consumption can improve end-to-end throughput.
Explicit frequency control therefore provides another way to return headroom to
application cores while retaining the service-core allocation.

\section{Outlook and Discussion}%
\label{sec:disc}

The fixed-budget model makes managing power and thermal headroom a
central concern for service-core systems. We now consider how this
perspective should shape network-stack policies and hardware interfaces,
and where its benefits depend on workload and platform constraints.

\textbf{\textit{Toward budget-aware network stacks.}}
Service-core designs should decide how much power and thermal
budget networking should consume while meeting latency objectives.
These decisions become more consequential as core counts grow faster
than package power.
Waiting policy, placement, and efficient core
reallocation~\cite{ousterhout:shenango,kaffes:shinjuku,fried:caladan}
can balance application throughput against network responsiveness,
with reallocation particularly useful for longer-term workload or locality
changes. On chiplet processors, these policies must account for headroom
at both die and package levels.
A key challenge is deciding when to retain, idle, slow, or
reallocate service cores as workload demand and package headroom change.
This reasoning extends to other systems, such as NFV,
where polling and compute-intensive applications can compete for package
headroom even on separate cores.

\textbf{\textit{Power-management interactions.}}
Established power-management techniques determine when and how our
findings apply. A tighter package power cap can induce contention at lower
concurrency; relaxing it can increase the benefit of additional active cores
if thermal headroom permits. Selective DVFS can release headroom while
preserving core ownership, as our RPC experiment illustrates
(\autoref{fig:motiv:network}), but excessive frequency reduction can make
networking the bottleneck. Idle-state selection similarly trades power
savings against wakeup latency. Released headroom improves throughput only
if productive cores can use it; frequency ceilings or memory bottlenecks
can limit the gain. Reallocation should be evaluated against
an efficiently waiting, power-managed baseline under the same power cap
and service-latency target. Comparisons should report application
throughput and service latency.

\textbf{\textit{The interface gap.}}
Hardware controls vary by platform. On our
EPYC~9655, the interfaces we tested did not let us explicitly clock selected
busy cores lower than others, although a single runtime parameter sufficed
on our older AMD and Intel systems. Our AWS experiment instead uses
frequency control over groups of 48 cores. Such differences constrain
software's allocation of the shared budget.

Waiting poses a related challenge. Network stacks largely rely
on busy polling that retains the core,
or blocking that relinquishes it to the scheduler.
Modern processors provide intermediate hardware mechanisms that return
increasing amounts of compute budget while preserving ownership of the
core, but these are coarse, difficult to use from userspace and lack adequate
portability. Service-core network stacks would
benefit from a standard ``halt-in-place'' primitive that separates
waiting from core ownership. It should let a service wait
for a queue update or deadline while retaining its core, with a
predictable wakeup cost.

\textbf{\textit{Limitations.}}
Our conclusions apply to processors operating near their package power
and thermal limits. Outside this regime, the classical model of
independent cores remains appropriate.
Our evaluation covers recent AMD platforms and focuses on throughput.
Evaluating other processor families, designing an adaptive policy, and
validating it against tail-latency targets under changing load remain
future work.
\section{Related Work}%
\label{sec:related}

\textbf{\textit{Waiting while retaining ownership.}}
Prior work analysed spinning versus blocking for
locks~\cite{karlin:competitive_spinning,boguslavsky:spinning_blocking},
and the energy-throughput trade-offs among spinning, \texttt{PAUSE},
\texttt{MWAIT}, and blocking~\cite{falsafi:unlocking_energy}.
Others reduce blocking and wakeup costs~\cite{humphries:context_switches},
exploit \texttt{MWAIT} for idle cores~\cite{baumann:barrelfish}, or
examine how virtualization hides hardware idleness~\cite{wang:mwatt_sched}.
We examine how these waiting choices affect the benefit of reclaiming
idle service cores under a shared power budget.

\textbf{\textit{Trading frequency for latency and energy.}}
A large body of work adapts frequency to workload needs, from sub-request DVFS
for latency-critical services~\cite{rubik,adrenaline} to application-controlled
scaling~\cite{hruby:slower_faster,dpdk_power,nicolas:dvfs_hpc_io}.
Other work uses feedback to control C-states~\cite{zhan:carb}.
The TURBO Diaries~\cite{wamhoff:turbo_diaries} explicitly studies slowing
or halting some cores to let others boost.
Race-to-idle designs such as
PowerNap~\cite{powernap} pursue energy proportionality~\cite{energy_proportionality}.
Our model shows that frequency reduction improves throughput without core reallocation.

\textbf{\textit{Reclaiming and harvesting idle capacity.}}
Microsecond schedulers reclaim idle service cores by
reallocation~\cite{fried:caladan,kaffes:shinjuku,ousterhout:shenango},
supported by low-overhead
handoffs~\cite{iyer:concord, aydogmus:xui,lin:fast_core_scheduling} and
policies for microsecond tasks~\cite{mcclure:efficient-scheduling}.
We qualify their shared premise that an idle core represents an
independent unit of compute. At the power cap, a reclaimed core runs at
reduced frequency, so recovered work is worth less than its core count
implies. An efficiently waiting core can already return much of its
budget without a software handoff.
Reallocation is worthwhile for long idle periods or below the power cap,
but not by default for workloads with microsecond idle periods under a
binding package power limit.
\section{Conclusion}

Modern processors increasingly behave as a shared compute budget
rather than a collection of independent cores. In this regime,
idle service cores are not necessarily wasted, and reclaiming
them at microsecond timescales can add complexity without
recovering meaningful compute.
We hope this perspective guides budget-aware network stacks 
on large multicore systems. \if \ANON 0
\section*{Acknowledgements}

We thank the anonymous reviewers and our shepherd, Gianni Antichi, 
for their helpful feedback. 
This work is supported in part by Deutsche Forschungsgemeinschaft
(DFG) --- 565608348, NSF grant 2450086, and the
University of Washington Center for the Future of Cloud Infrastructure
(FOCI).
 \fi

\bibliographystyle{plain}
\bibliography{paper,bibdb/strings,bibdb/papers,bibdb/defs}

\label{page:last}
\end{document}